\documentclass[onecolumn,floatfix,superscriptaddress,a4paper,
               showpacs,showkeys,nofootinbib,preprint]{revtex4}

\usepackage{epsfig}
\usepackage{latexsym}
\usepackage{xspace}
\usepackage{hyperref}
\usepackage[latin2]{inputenc}
\usepackage{indentfirst}
\usepackage{enumerate}

\usepackage{amsmath}
\usepackage{amssymb}
\usepackage[english]{babel}
\usepackage{url}

\newcommand{\bs}{\boldsymbol}

\newcommand*{\dis}{\displaystyle}

\begin{document}

\title{\bf Phase Diagrams of Relativistic Selfinteracting Boson System
}
\author{V. Gnatovskyy}
\affiliation{Bogolyubov Institute for Theoretical Physics, 03143 Kyiv, Ukraine}
\author{D. Anchishkin}
\affiliation{Bogolyubov Institute for Theoretical Physics, 03143 Kyiv, Ukraine}
\affiliation{Taras Shevchenko National University of Kyiv, 03022 Kyiv, Ukraine}
\affiliation{Frankfurt Institute for Advanced Studies, Ruth-Moufang-Strasse 1,
60438 Frankfurt am Main, Germany}
\author{D. Zhuravel}
\affiliation{Bogolyubov Institute for Theoretical Physics, 03143 Kyiv, Ukraine}
\author{V. Karpenko}
\affiliation{Taras Shevchenko National University of Kyiv, 03022 Kyiv, Ukraine}



\begin{abstract}
Within the Canonical Ensemble, we investigate a system of
interacting relativistic bosons at finite temperatures and finite
isospin densities in a mean-field approach.
The mean field contains both attractive and repulsive terms.
Temperature and isospin-density dependencies of thermodynamic quantities
were obtained.
It is shown that in the case of attraction between particles in a bosonic
system, a liquid-gas phase transition develops against the background of
the Bose-Einstein condensate.
The corresponding phase diagrams are given.
We explain the reasons why the presence of a Bose condensate significantly
increases the critical temperature of the liquid-gas phase transition compared
to that obtained for the same system within the framework of the Boltzmann
statistics.
Our results may have implications for the interpretation of experimental data,
in particular, how sensitive the critical point of the mixed phase is to the
presence of the Bose-Einstein condensate.

\end{abstract}

\maketitle

\section{Introduction}
\label{sec1}
It is commonly accepted that QCD exhibits a rich phase structure at
finite temperatures and baryon densities, for instance, the transition
from hadron gas to quark-gluon plasma, the transition from chiral
symmetry breaking to the symmetry restoration \cite{bzdak-esum-2020}.
The physical motivation to study QCD at finite isospin density and the
corresponding pion system is related to the investigation
of compact stars, isospin asymmetric nuclear matter and heavy ion collisions.
In early studies of dense nuclear matter and compact stars, it has been suggested
that charged pions and even kaons are condensed at sufficiently high densities.
The knowledge of the phase structure of the meson systems, in the regime
of finite temperatures and isospin densities, is crucial for understanding
a wide range of phenomena from nucleus-nucleus collisions to boson, neutron
stars, and cosmology.
This field is essential to investigations of the hot and dense hadronic matter,
a subject of active research.
Meanwhile, investigations of the meson systems have their specifics due to the
possibility of the Bose-Einstein condensation (BEC) of the boson particles.
Formation of classical pion fields in heavy-ion collisions was discussed in
Refs.~\cite{anselm-1991,blaizot-1992,bjorken-1992,mishustin-greiner-1993}.
Then, the study on the QCD phase structure is extended to finite isospin
densities and the systems of pions and K-mesons  with a finite isospin
chemical potential have been considered in more recent studies
\cite{son-2001,kogut-2001,toublan-2001,mammarella-2015,carignano-2017,mannarelli-2019}.
First-principles lattice  calculations provide a solid basis for our
knowledge of the finite temperature regime.
New results concerning dense pion systems have been obtained
recently using lattice methods \cite{brandt-2016,brandt-2017,brandt-2018}.

In the present paper we consider an interacting particle-antiparticle
boson system at the conserved isospin (charge) density $n_I$ and finite
temperature $T$.
We name the bosonic particles as ``pions'' just conventionally.
The preference is made because the charged $\pi$-mesons are the lightest hadrons
that couple to the isospin chemical potential.
On the other hand, the pions are the lightest nuclear boson particles and thus,
an account for ``temperature creation'' of particle-antiparticle pairs is a
relevant problem based on quantum statistics.
To account for the interaction between the bosons we introduce a
phenomenological Skyrme-like mean field $U(n)$, which depends only
on the total meson density $n$.
We regard such a selfinteracting many-particle system as a
toy-model that can help us understand the BEC and
phase transitions over a wide range of temperatures and densities.
The mean field $U(n)$ rather reflects the presence of other strongly interacting
particles in the system, for instance $\rho$-mesons and nucleon-antinucleon
pairs at low temperatures or gluons and quark-antiquark pairs at high
temperatures, $T > T_{\rm qgp} \approx 160$~MeV.

The presented study is a part of a sequel
\cite{anchishkin-mishustin-2019,mishustin-anchishkin-2019,
anchishkin-4-2019,stashko-anchishkin-2020} that started with the investigation
of an interacting particle-antiparticle boson system at $\mu_I = 0$.
The next development of the subject was given
in Ref.~\cite{anch-2022-prc}, where the boson system
was considered within the framework of the Canonical Ensemble with the
canonical variables $(T,\, n_I)$, i.e. at the conserved isospin (charge) density.
In this formulation in \cite{anch-2022-prc} we calculated the temperature
characteristics of a non-ideal hot ``pion'' gas with a fixed isospin density
$n_I = n^{(-)} - n^{(+)} > 0$, where $n^{(\mp)}$ are the particle-number
densities of the $\pi^{-}$ and $\pi^{+}$ mesons, respectively.
In the present study we proceed to exploit the Canonical Ensemble.
But now, we focus on the isospin-density dependencies of thermodynamic quantities
when the temperature is fixed.

\section{The mean-field model for the system  \newline
of particles and antiparticles}
\label{sec:mfm-part-antipart}
Our consideration of thermodynamic properties of the system of
interacting bosonic particles and antiparticles at finite temperatures
is carried out within the framework of the thermodynamic mean-field
model, which was introduced in Refs.~\cite{anch-1992,anchsu-1995}
and further developed in Ref.~\cite{anch-vovchenko-2015}.
This approach is based on the representation of the free energy $F$ of the
particle-antiparticle system as the sum of two parts: the first part $F_0$ is
the free energy of the two-component system of free particles, and the second
part $F_{\rm int}$ is responsible for the interaction between all particles,
i.e., $F = F_0 + F_{\rm int}$.
Therefore, it is assumed that in general the free-energy density of the
two-component system looks like
\begin{equation}
\phi\left(T,n_1,n_2\right)\,=\, \phi^{(0)}_1\left(T,n_1\right)
+ \phi^{(0)}_2 \left(T,n_2\right) + \phi_{\rm int}\left(T,n\right) \,,
\label{eq:phi-structure}
\end{equation}
where $\phi = F/V$ with $V$ as the volume of the system,
$\phi^{\left(0\right)}_1$ and $\phi^{\left(0\right)}_2$ are the free
energy densities for the free particles of the first and second components,
respectively,
whereas the density of free energy $\phi_{\rm int}$ takes into account the
interaction in the system, $n_1$ and $n_2$ is the particle-number density
of each component and $n = n_1 + n_2$ is the total particle-number density.
Next, the chemical potential associated with every component is calculated
as correspondent derivative
%
\begin{equation}
\mu_i = \left[\frac{\partial \phi(T,n_1,n_2)}{\partial n_i}\right]_T \,,
\label{eq:chem-pot}
\end{equation}
where $i = 1,\,2$.
This results in
\begin{equation}
\mu^{(0)}_i = \mu_i(T,n_i) - U(T,n) \,,
\label{eq:mu0}
\end{equation}
where we define
\begin{eqnarray}
\mu^{(0)}_i \,=\, \frac{\partial \phi^{(0)}_i(T,n_i)}{\partial n_i} \,,
\qquad
U(T,n)  \, \equiv \, \frac{\partial \phi_{\rm int}(T,n)}{\partial n} \,.
\label{eq:def-mu0-u}
\end{eqnarray}
Similarly, one can write the pressure, $p = \mu_1 n_1 + \mu_2 n_2 - \phi$,
dividing it into free and interacting parts
\begin{eqnarray}
p\left(T,n_1,n_2\right) \,=\, p^{(0)}_1 + p^{(0)}_2 + P_{\rm ex}(T,n) \,,
\label{eq:p-total}
\end{eqnarray}
where $p^{(0)}_i = \mu^{(0)}_i n_i - \phi^{(0)}_i$ is the pressure of the
ideal gas created by the $i$-th component of the system and
\begin{equation}
P_{\rm ex}(T,n) \, \equiv \,
n\, \left[\frac{\partial \phi_{\rm int}(T,n)}{\partial n}\right]_T \,-\, \phi_{\rm {int}} \,,
\label{eq:def-pex}
\end{equation}
is the excess pressure.
It is seen that the definitions of $U(T,n)$ and $P_{\rm ex}(T,n)$ lead to a
differential correspondence between these quantities:
\begin{equation}
n\, \left[\frac{\partial U(T,n)}{\partial n}\right]_T \,
=\, \left[\frac{\partial P_{\rm ex}(T,n)}{\partial n}\right]_T \,.
\label{eq:dif-relation}
\end{equation}

We limit our consideration to the case when at a fixed temperature
the interacting boson particles and boson antiparticles are in dynamic
equilibrium with respect to the processes of annihilation and pair-creation.
Due to the opposite sign of the charge, the chemical potentials of the bosonic
particles $\mu_1$ and the bosonic antiparticles $\mu_2$ have opposite signs
(for details, see \cite{anch-vovchenko-2015}):
\begin{equation}
\mu_1 \,=\, -\mu_2 \, \equiv \, \mu_I \,.
\label{eq:23}
\end{equation}
Therefore, the Euler relation includes only the isospin number density,
$n_{I} = n^{(-)} - n^{(+)}$, in the following way
\begin{equation}
\varepsilon \,+\, p \,=\, T\,s \,+\, \mu_I \, n_{I} \,,
\label{eq:24}
\end{equation}
where $n_1 \to n^{(-)}$ is the particle-number density of bosonic particles, and
$n_2 \to n^{(+)}$ is the particle-number density of bosonic antiparticles,
$\varepsilon$ is the energy density and $s$ is the entropy density.
\footnote{We use the negative total electric charge in the system because of
the predominance of the creation of negative pions over positive ones in
relativistic nucleus-nucleus collisions.  }
In what follows, we will consider the boson particle-antiparticle
system with conserved isospin number density $n_I$,
whereas in this study the total particle-number density $n$
is a thermodynamic quantity that depends on $T$ and $n_I$, i.e., $n(T,n_I)$.
\footnote{The dynamical conservation of the total number of pions in a pion-enriched
system created on an intermediate stage of a heavy-ion collision was considered
in Refs.~\cite{kolomeitsev-voskresensky-2018,kolomeitsev-borisov-voskresensky-2018,
kolomeitsev-voskresensky-2019}
}

Exploiting Eq.~(\ref{eq:p-total}) with formula for ideal gas in
the Grand Canonical Ensemble the total pressure in the particle-antiparticle
system reads
\footnote{Here and below we adopt
the system of units $\hbar=c=1$, $k_{_B}=1$}
%
\begin{eqnarray}
p &=& -\,  T \int  \frac{d^3k}{(2\pi)^3} \,
\ln{ \left[ 1 - \exp \left( -\frac{\sqrt{m^2+{\bf k}^2} +  U(T,n) - \mu_I }{T}\right)
\right] } \, -
\nonumber  \\
&& \hspace{1mm}
-\,  T \int  \frac{d^3k}{(2\pi)^3} \,
\ln{ \left[ 1 - \exp \left( -\frac{\sqrt{m^2+{\bf k}^2} + U(T,n) + \mu_I }{T}\right) \right] } \,
+\, P_{\rm ex}(T,n) \,,
\label{eq:d16}
\end{eqnarray}
where $\mu^{(0)}_1$ and $\mu^{(0)}_2$ are altered for $(\mu_I - U)$ and
$(-\mu_I - U)$, respectively, in accordance with Eq.~(\ref{eq:mu0}) and
Eq.~(\ref{eq:23}).

The thermodynamic consistency of the mean-field model can be obtained by putting
in correspondence of two expressions which  must coincide in the result.
These expressions, which determine the isospin density, read
\begin{equation}
n_{I} \,=\, \left(\frac{\partial p}{\partial \mu_I}\right)_T \,,
\quad {\rm and} \quad
n_{I} \,=\,  \int \frac{d^3k}{(2\pi )^3} \,
\big[ f_{_{\rm BE}}\big(E(k,n),\mu_I\big) \,-\, f_{_{\rm BE}}\big(E(k,n),-\mu_I\big)\big] \,,
\label{eq:i-density}
\end{equation}
where pressure is given by Eq.(\ref{eq:d16}).
Here $E(k,n) = \omega_k + U(T,n)$ with $\omega_k = \sqrt{m^2 + {\bf k}^2}$ and
the Bose-Einstein distribution function reads
\begin{equation}
f_{_{\rm BE}}\big(E,\mu\big) \,
=\, \left[ \exp{ \left( \frac{E - \mu}{T} \right)}  - 1\right]^{-1} \,.
\label{eq:32}
\end{equation}
In order the expressions (\ref{eq:i-density})  
to coincide in the result the following relation between the mean field and
the excess pressure arises as the necessary condition
\begin{equation}
n\, \frac{\partial U(T,n)}{\partial n} \,=\, \frac{\partial P_{\rm ex}(T,n)}{\partial n} \,.
\label{eq:d20}
\end{equation}
As we see this relation coincides literally with relation (\ref{eq:dif-relation}),
which was derived using definitions of the mean field $U(T,n)$ and excess
pressure $P_{\rm ex}(T,n)$ in Eq.~(\ref{eq:def-mu0-u}) and in Eq.~(\ref{eq:def-pex}),
respectively.
The relation (\ref{eq:d20}) that provides the thermodynamic consistency of the
model has a natural basis because there is only one source for both quantities
$U$ and $P_{\rm ex}$, it is interaction in the system.

\subsection{Parametrization of the mean field }
\label{sec:skyrme-param}

The thermodynamic mean-field model has been applied for several
physically interesting systems including the hadron-resonance gas
\cite{anch-vovchenko-2015} and the pionic gas \cite{anch-2016}.
This approach was extended to the case of a bosonic system at $\mu_I = 0$ which
can undergo Bose condensation
\cite{anchishkin-mishustin-2019, anchishkin-4-2019}.
In the present study a generalized formalism given in section
\ref{sec:mfm-part-antipart} is used to describe the particle-antiparticle system
of bosons when the  isospin density is finite, $n_I \neq 0$.
At the same time, we assume that the interaction between particles is described
by the Skyrme-like mean field which depends on the total particle-number
density $n$.
The latter means that we take into account just a strong interaction.
For further calculations we adopt the following form of the mean field
%
\begin{equation}
U(n)\, =\, - \, A\,n\, +\, B\, n^2 \,,
\label{eq:mf1}
\end{equation}
where $A$ and $B$ are the model parameters, which should be specified.
In accordance with relation (\ref{eq:d20}) we calculate the excess pressure
 \begin{equation}
P_{\rm ex}(n) \,=\, - \, \frac 12 A\,n^2\, +\, \frac 23 B\, n^3 \,,
\label{eq:pex-skyrme}
 \end{equation}
The mean field $U(n)$ can be thought as some effective one which includes several
contributions.
For instance, investigation of the properties of
a dense and hot pion gas is well inspired by formation of the medium with low
baryon numbers at midrapidity what was proved in the experiments
at RHIC and LHC \cite{adamczyk-2017,abelev-2012}.
By this reasons, in our calculations we consider a general case of $A > 0$,
to study a bosonic system with both attractive and repulsive contributions to
the mean field (\ref{eq:mf1}).
Our main goal is a study how the relation between repulsion and attraction
in the system influences on the Bose-condensation and thermodynamic properties
of the bosonic system.
In the present paper to investigate these features we keep the repulsive
coefficient $B$ as a constant, whereas the coefficient $A$, which determines
the intensity of attraction of the mean field (\ref{eq:mf1}), will be varied.
%
To do this it is advisable to parameterize the coefficient $A$
with the help of solutions of equation
$U(n) + m = 0$, similar to parametrization adopted in
Refs.~\cite{anchishkin-mishustin-2019,anchishkin-4-2019}.
For the given mean field (\ref{eq:mf1}) there are two roots of this
equation ($n_{1,2} = (A \mp \sqrt{A^2 - 4mB})/2B$)
\begin{equation}
n_1 \, =\, \sqrt{\frac{m}{B}} \left( \kappa - \sqrt{\kappa^2 - 1}\right)\,,
\qquad
n_2 \, =\, \sqrt{\frac{m}{B}} \left( \kappa + \sqrt{\kappa^2 - 1} \right) \,,
\quad {\rm where}  \quad
\kappa \, \equiv\, \frac{A}{2\,\sqrt{m \,B}} \,.
\label{eq:n1-n2-10}
\end{equation}
Then, one can parameterize the attraction coefficient as $A = \kappa A_{\rm c}$
with  $A_{\rm c} = 2\sqrt{m B}$.
As we will show below, the dimensionless parameter $\kappa$ is the scale parameter
of the model
which determines the phase structure of the system.
As it is seen from Eq.~(\ref{eq:n1-n2-10}) for the values of parameter $\kappa < 1$
there are no real roots.
The critical value  $A_{\rm c}$ is obtained when both roots coincide, i.e.
when $\kappa = 1$, then $A = A_{\rm c} = 2\sqrt{m B}$.

We consider two intervals of the parameter $\kappa$.
First interval corresponds to  $\kappa \le 1$,  there are no real roots of
equation $U(n) + m = 0$.
We associate these values of $\kappa$ with a ``weak'' attractive interaction
and in the present study we consider variations in the attraction coefficient $A$
for values of $\kappa$ only from this interval.
Second interval corresponds to  $\kappa > 1$,  there are two real roots
of equation $U(n) + m = 0$.
We associate this interval with a ``strong'' attractive interaction.
This case will be considered elsewhere.

\section{The phase transition to the Bose-Einstein condensate}
\label{sec:bec-phase-transition}

In the mean-field approach the behavior of the particle-antiparticle boson
system, when both components in thermal (kinetic) phase, is determined by the
set of two transcendental equations
%
\begin{eqnarray}
\label{eq:tot-n}
n &=&  \int \frac{d^3k}{(2\pi )^3} \,
\left[ f_{_{\rm BE}}\big(E(k,n),\mu_I\big) \,
+\, f_{_{\rm BE}}\big(E(k,n),-\mu_I\big) \right] \,,
\\
n_{I} &=&  \int \frac{d^3k}{(2\pi )^3} \,
\left[ f_{_{\rm BE}}\big(E(k,n),\mu_I\big) \,
-\, f_{_{\rm BE}}\big(E(k,n),-\mu_I\big) \right] \,,
\label{eq:ni}
\end{eqnarray}
where the Bose-Einstein distribution function $f_{_{\rm BE}}\big(E,\mu_I\big)$ is
defined in (\ref{eq:32}) and $E(k,n) =  \omega_k + U(n)$, the degeneracy factor
$g = 1$ because the spin of particles is zero.
Equations (\ref{eq:tot-n})-(\ref{eq:ni}) should be solved selfconsistently with
respect to $n$ and $\mu_I$  for the given canonical variables $(T,\, n_I)$.
In the present we consider boson system in the Canonical Ensemble.
In this approach the chemical potential $\mu_I$ is a thermodynamic quantity which
depends on the canonical variables, i.e. $\mu_I(T,n_I)$.

In case of the cross state, when the particles, i.e. $\pi^-$-mesons, are in the
condensate phase and antiparticles are still in the thermal (kinetic) phase,
Eqs.~(\ref{eq:tot-n}), (\ref{eq:ni}) should be generalized to
include condensate component $n^{(-)}_{\rm cond}$.
We should take into account also that the particles ($\pi^-$ or
high-density component) can be in condensed state just under the
necessary condition
\begin{equation}
U(n) \,-\, \mu_I  \,=\,  - m \,.
\label{eq:condens-cond}
\end{equation}
During decreasing of temperature from high values, where both
$\pi^-$ and $\pi^+$ are in the thermal phase, the density of $\pi^-$-component
$n^{(-)}(T,n_I)$ achieves first the critical curve at temperature
$T_{\rm cd}$, where at the crossing point condition (\ref{eq:condens-cond}) is
valid, but the density of the condensate is zero at this point,
i.e., $n_{\rm cond} = 0$.
This means that the curve $n_{\rm lim}(T)$, which is defined as
\begin{equation}
n_{\rm lim}(T)  \,=\,  \int \frac{d^3k}{(2\pi)^3}\,
f_{_{\rm BE}}\big(\omega_k,\mu_I\big)\Big|_{\mu_I = m} \,,
\label{eq:nlim-id}
\end{equation}
is  the critical curve for $\pi^-$-mesons or for high-density
component of the gas.
As we see function (\ref{eq:nlim-id}) represents the maximal density of thermal
(kinetic) boson particles of the ideal gas at temperatures $T \le T_{\rm cd}$
because the chemical potential has its maximum allowed value.
Hence, we obtain that the critical curve of the particle-antiparticle boson
system calculated in the mean-field approach coincides with the critical curve
for the ideal gas.

With account for Eqs.~(\ref{eq:condens-cond}) and (\ref{eq:nlim-id}) we write
the generalization of the set of Eqs.~(\ref{eq:tot-n}), (\ref{eq:ni})
\begin{eqnarray}
\label{eq:tot-n2a}
n &=&  n^{(-)}_{\rm cond}(T) + n_{\rm lim}(T) \,
+\, \int \frac{d^3k}{(2\pi )^3} \,f_{_{\rm BE}}\big( E(k,n),-\mu_I \big) \,,
\\
n_{I} &=&  n^{(-)}_{\rm cond}(T) + n_{\rm lim}(T)
- \int \frac{d^3k}{(2\pi )^3} \, f_{_{\rm BE}}\big(E(k,n),-\mu_I\big)  \,,
\label{eq:ni2a}
\end{eqnarray}
where $\mu_I = m + U(n)$.
One can see from Eqs.~(\ref{eq:tot-n2a}), (\ref{eq:ni2a}) that the
particle-number density $n^{(+)}$ is provided only by thermal $\pi^{+}$ mesons.
Whereas, the density $n^{(-)}$ is provided by two fractions:
the  condensed particles ($\pi^{-}$ mesons at $\bs k = 0$) with the
particle-number density $n^{(-)}_{\rm cond}(T)$, and
thermal $\pi^{-}$ mesons  at $|\bs k| > 0$ with the particle-number density
$n_{\rm lim}(T)$.
Hence, the particle-density sum rule for the phase of $\pi^{-}$ mesons only
in the interval $T < T_{\rm cd}$ reads: $n^{(-)} = n^{(-)}_{\rm cond}(T) + n_{\rm lim}(T)$.
The set of equations (\ref{eq:tot-n2a}) - (\ref{eq:ni2a}) can be reduced to one
equation with respect to the total number density $n$:
\begin{equation}
n  \,=\, n_I \,+\, 2 \int \frac{d^3k}{(2\pi )^3} \,
f_{_{\rm BE}}\big( E(k,n),-\mu_I \big)\big|_{\mu_I = m + U(n)} \,.
\label{eq:ntot-cond}
\end{equation}

It is necessary to note, that expression ``particles are in the condensate phase''
is, of course, a conventional one, because in the essence it is a mixture phase,
where at a fixed temperature some fraction of particles, i.e. a fraction
of $\pi^-$-mesons, belongs to thermal phase with momentum $|\bs k| > 0$ and other
fraction of  $\pi^-$-mesons belongs to the Bose-Einstein condensate, where
all $\pi^-$-mesons have zero momentum, $\bs k = 0$.

\medskip

For further evaluation of the phase diagram, it is necessary to know the
pressure value.
First, we calculate the pressure for the state of the system when
$\pi^-$ and $\pi^+$ mesons are in the thermal (kinetic) phase
\begin{equation}
p \,=\, \frac{1}{3} \int \frac{d^3k}{(2\pi)^3}\,\frac{{\bf k}^2}{\omega_k}
\big[ f_{_{\rm BE}}\big(E(k,n),\mu_I\big) \,
+\, f_{_{\rm BE}}\big(E(k,n),-\mu_I\big)\big] \,+\, P_{\rm ex}(n) \,,
\label{eq:30-2}
\end{equation}
where $n(T,n_I)$ and $\mu_I(T,n_I)$ are solution of Eqs.~(\ref{eq:tot-n}),
(\ref{eq:ni}).

\begin{figure}
\centering
\includegraphics[width=0.49\textwidth]{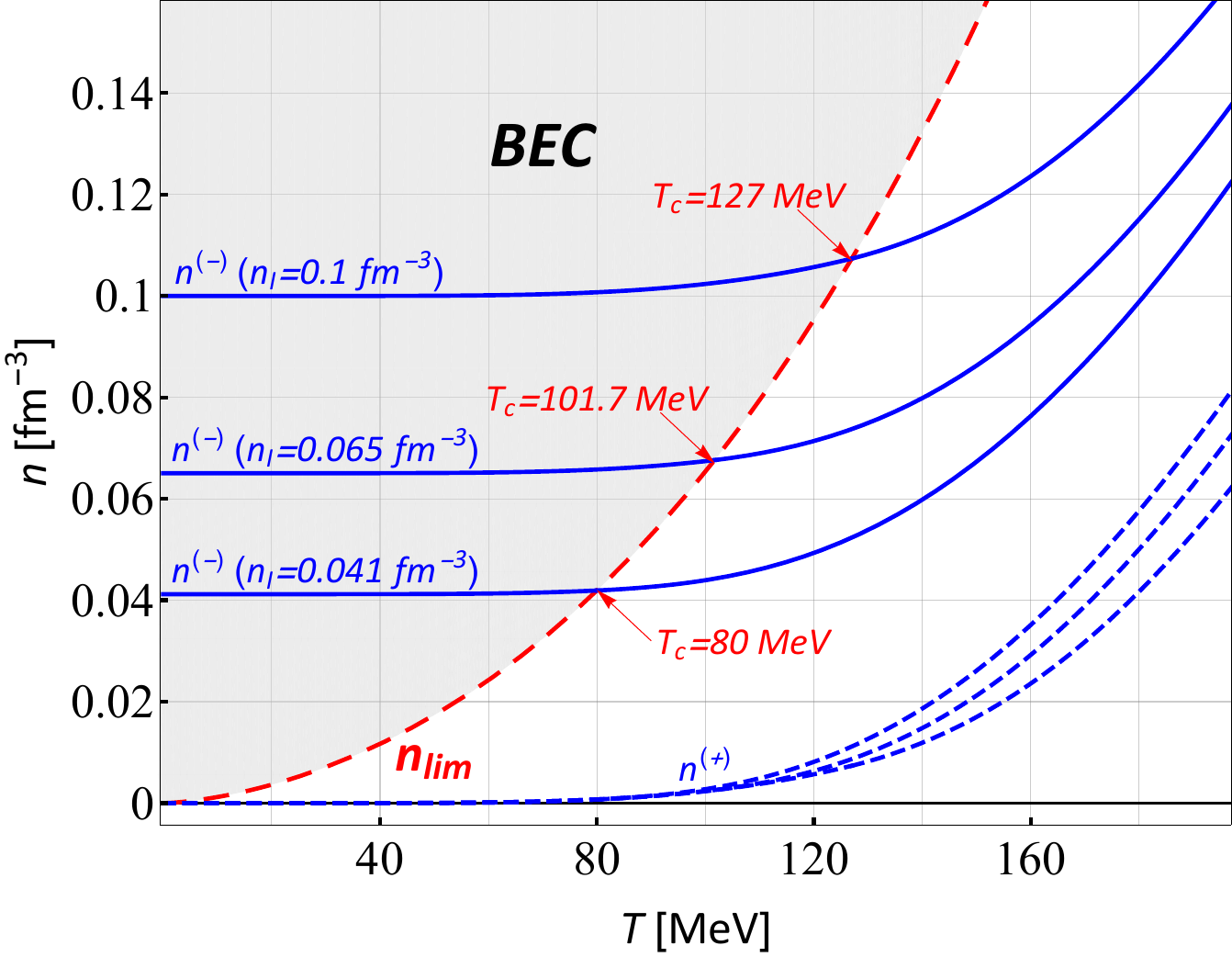}
\caption{
Temperature dependence of the particle-number density of $\pi^-$ mesons,
$n^{(-)}$ (blue solid lines), and $\pi^+$ mesons, $n^{(+)}$
(blue dashed lines), for the interacting $\pi^+$-$\pi^-$ pion gas at
$\kappa = 0.1$ and $n_I = 0.041,\, 0.065,\, 0.1$~fm$^{-3}$.
The red dashed line is the critical curve $n_{\rm lim}$ that captures the
particle density of a single-component ideal gas at $\mu = m$.
The shaded area, labeled BEC, indicates the region where the $\pi^-$ mesons
form a Bose-Einstein condensate.
In the figure, the temperature $T_{\rm cd}$ is denoted as $T_{\rm c}$.
}
\label{fig:article-n-vs-T}
\end{figure}

As we mentioned earlier, we consider the ``weak'' attraction between particles,
i.e., $\kappa \le 1$.
As was shown in \cite{anch-2022-prc} for this interval of the attraction parameter,
a Bose-Einstein condensate can be developed only for the dominant component of
the particle-antiparticle system; in our case, only $\pi^-$ mesons can be in the
condensate phase at low temperatures
\footnote{When $\kappa > 1$, both components of the particle-antiparticle system
can be in the condensate phase at low temperatures.}.
That is, taking into account Eqs.~(\ref{eq:condens-cond}),
for the condensate phase $\pi^-$ mesons and the thermal phase $\pi^+$ mesons,
we calculate the pressure as
\begin{equation}
p = \frac{1}{3} \int \frac{d^3k}{(2\pi)^3}\,\frac{{\bf k}^2}{\omega_k}
\big[ f_{_{\rm BE}}\big(\omega_k, m\big) \,
+\, f_{_{\rm BE}}\big(E(k,n),-\mu_I\big)\big|_{\mu_I = U(n) + m}\big] \,
+\, P_{\rm ex}(n) \,,
\label{eq:30-4}
\end{equation}
where $\mu_I = U(n) + m$ and with this value of the chemical potential for the
argument of the first distribution function on the r.h.s. of this equation, we
get $E(k,n) - \mu_I = \sqrt{m^2 + \bs k^2} - m$.
The total number density $n(T,n_I)$ used in (\ref{eq:30-4}) is
a solution to eq.~(\ref{eq:ntot-cond}).
That is, for visual interpretation, it can be said that eq.~(\ref{eq:30-4})
``determines the pressure on the curve $n(T,n_I)$'';  we can see one of such
curves, for example, in Fig.~\ref{fig:article-n-vs-T} for a fixed value $n_I$.
Note that the condensed particles, i.e. the fraction of $\pi^-$ mesons that are
in the condensate, do not contribute to the pressure.
This feature of pressure significantly distinguishes the bosonic system from
systems without condensate and is decisive for the liquid-gas phase transition
and the value of the critical point of the phase diagram.

\subsection{Thermodynamic quantities}
\label{sec:numeric-calc}

In what follows a value of the repulsive coefficient $B$
of the mean field (\ref{eq:mf1}), is fixed.
At the same time the coefficient $A$, which determines the intensity of
attraction of the mean field (\ref{eq:mf1}), will be varied.
The coefficient $B$
is obtained from an estimate based on the virial expansion \cite{hansen-2005},
$B = 10 m v_0^2$ with $v_0$ equal to four times the proper volume of
a particle, i.e. $v_0 = 16 \pi r_0^3/3$.
We take $v_0 = 0.45$~fm$^3$ that corresponds to a ``particle radius''
$r_0 \approx 0.3$~fm.
The numerical calculations will be done for bosons with mass $m = 139$~MeV,
which we call ``pions''.
In this case the repulsive coefficient is $B/m = 2.025$~fm$^6$ and it is kept
constant through all present calculations.

At high temperatures, i.e. $T \ge T_{\rm cd}$, both components of the
particle-antiparticle boson system are in the thermal phase,
and thermodynamic properties of the system are determined by a set of
equations (\ref{eq:tot-n}) and (\ref{eq:ni}).
Solving this set for given values of $T$ and $n_I$, we obtain the functions
$\mu_I(T,n_I)$ and $n(T,n_I)$ and then it is possible to calculate
other thermodynamic quantities.

With decreasing temperature, the particle density $n^{(-)}(T)$ crosses the
critical curve in the point, which corresponds to the value $T = T_{\rm cd}$,
see Fig.~\ref{fig:article-n-vs-T} (on the graph, $T_{\rm cd}$ is denoted
as $T_{\rm c}$).
The dependence of $\pi^-$-meson density on temperature is depicted as blue solid
line, the dependence of $\pi^+$-meson density as blue dashed line.

During further decreasing of temperature in the interval $T < T_{\rm cd}$
the $\pi^{-}$-mesons start to ``drop down'' into the condensate state,
which is characterized by the value of momentum $\bs k = 0$.
In the limit, when $T = 0$, all particles of the high-density component,
i.e. $\pi^{-}$ mesons, will be in the condensate phase.
At the same time, the density of the low-density component or $\pi^{+}$ mesons,
which are in the thermal phase, decreases with decreasing temperature and it
becomes zero at $T = 0$.
For the temperatures less than the critical one, i.e. $T < T_{\rm cd}$,
the thermodynamic properties of the system are determined by
Eqs.~(\ref{eq:tot-n2a}), (\ref{eq:ni2a}), where we take into account that
$\mu_I = U(n) + m$ for all temperatures of this interval unless the high-density
component $n^{(-)}$ is in condensed state
\footnote{We apply our consideration to the pion gas with $n_I = n^{(-)} - n^{(+)} > 0$.}.

\begin{figure}[h]
\centering
\includegraphics[width=0.49\textwidth]{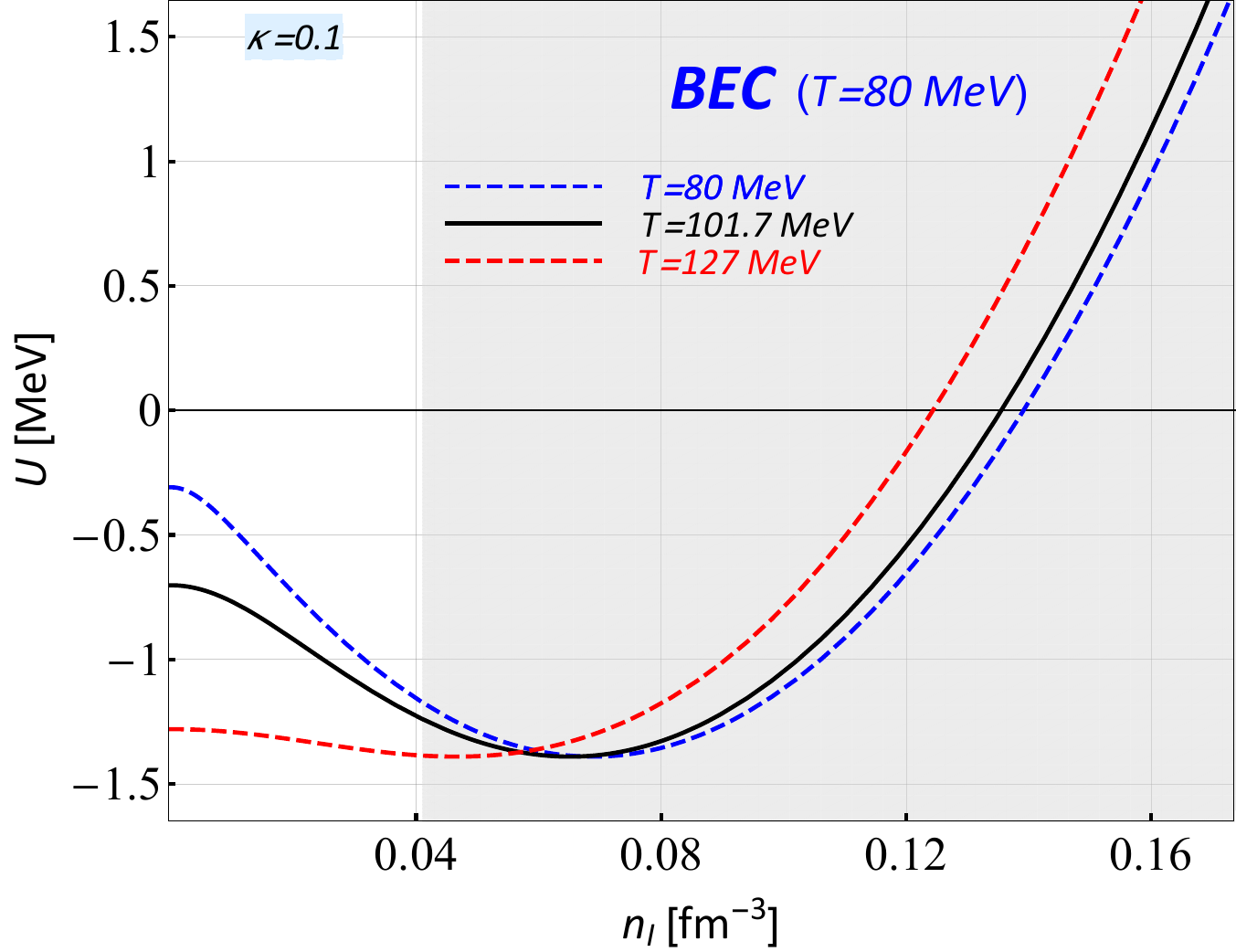}
\includegraphics[width=0.49\textwidth]{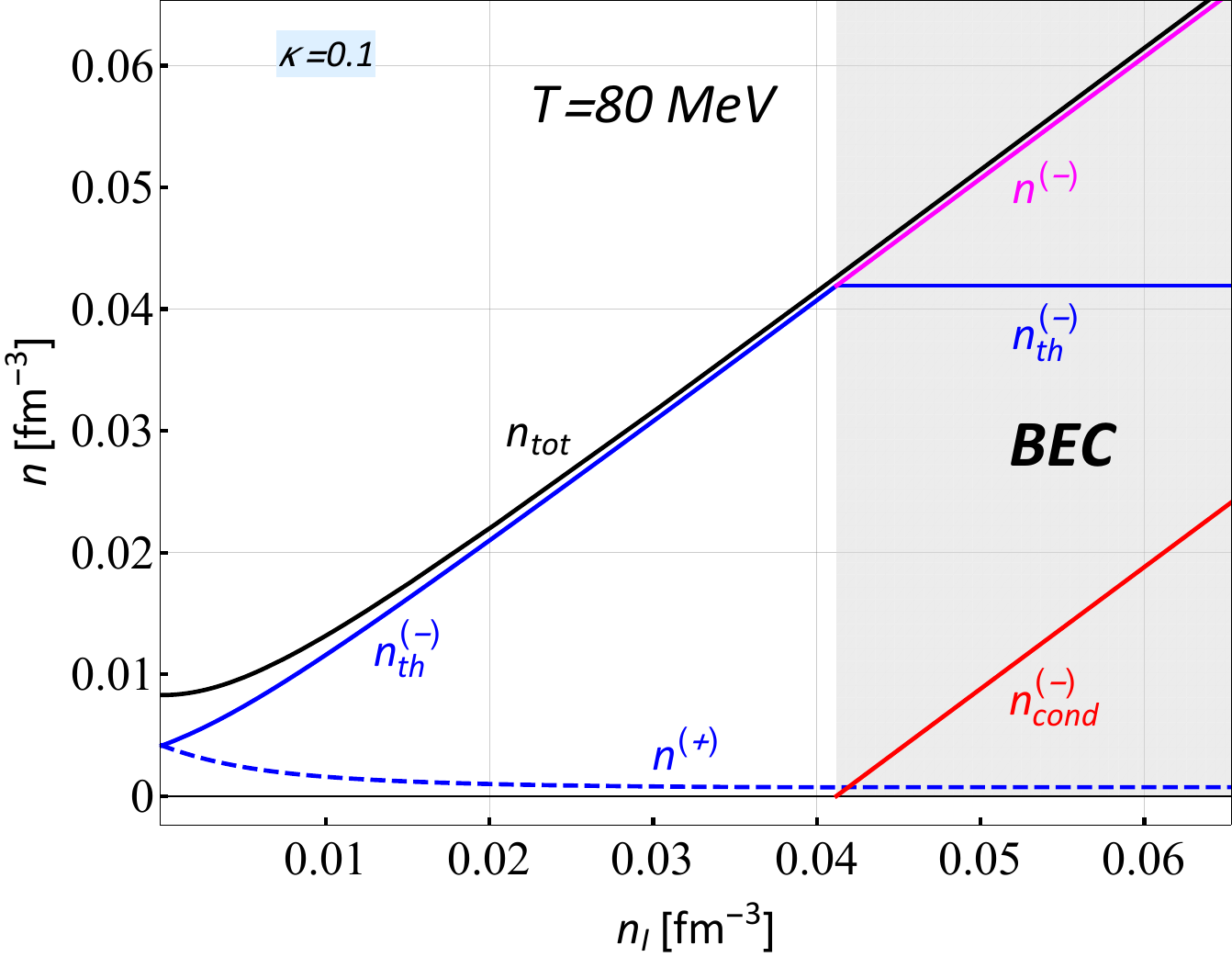}
\caption{ In both panels, the dashed area labeled as BEC represents the
condensate states of the interacting $\pi^+$-$\pi^-$ pion gas in the
mean-field model.
{\it Left panel:} Dependence of the mean field $U(n)$ on the isospin density
$n_I$ at $\kappa = 0.1$ and $T = 80,\, 101.7,\, 127$~MeV.
{\it Right panel:} The particle-number densities $n^{(+)}$, $n^{(-)}$ and
$n_{\rm tot} = n^{(+)} + n^{(-)}$ against the isospin density $n_I$
at $T = 80$~MeV and $\kappa = 0.1$.
Here $n^{(-)}_{\rm th}$ and $n^{(-)}_{\rm cond}$ are the particle density of
thermal and condensed $\pi^-$ mesons, respectively.
}
\label{fig:n-vs-ni-k01}
\end{figure}
%

Equation (\ref{eq:tot-n2a}) can be used to determine the critical
temperature $T_{\rm cd}$.
Indeed, let us take into account that at the point of intersection with the
critical curve, the condensate density is zero,
$n^{(-)}_{\rm cond}\big(T_{\rm cd}\big) = 0$,
and the density of thermal $\pi^-$ mesons becomes equal to
$n^{(-)}\big(T_{\rm cd}\big) = n_{\rm lim}\big(T_{\rm cd}\big)$.
Then, at the temperature $T_{\rm cd}$ on the l.h.s. of Eq.~(\ref{eq:tot-n2a})
we have $n = 2 n_{\rm lim}\big(T_{\rm cd}\big) - n_I$.
Therefore, at this temperature point on the critical curve
Eq.~(\ref{eq:tot-n2a}) reads
\begin{equation}
n_{\rm lim}(T) - n_I = \int \frac{d^3k}{(2\pi )^3} \,f_{_{\rm BE}}\big( E(k,n),-\mu_I
\big)\Big|_{\mu_I = U(n) + m}
\quad {\rm with}  \quad
E(k,n) = \omega_k + U\!\left( 2n_{\rm lim}(T) - n_I \right) \,.
\label{eq:nneg-ST2-2}
\end{equation}
Solving Eq.~(\ref{eq:nneg-ST2-2}) with respect to $T$ at $\kappa = 0.1$
for $n_I = 0.041,\, 0.065,\, 0.1$~fm$^{-3}$, we obtained the transition
temperatures $T_{\rm cd} = 80,\, 102,\, 127$~MeV, respectively
(see Fig.~\ref{fig:article-n-vs-T}).

Hence, it turns out that the temperature $T_{\rm cd}^{(-)}$ determines the phase
transition to BEC for whole pion system because the antiparticles ($\pi^+$-mesons)
are completely in thermal state for all temperatures and thus, the condensate
is created just by the particles of high-density component $n^{(-)}(T)$.
Then, the total density of condensate in the two-component pion system
at ``weak'' attraction, i.e. at $\kappa \le 1$,
is created by $\pi^-$-mesons only, i.e. $n_{\rm cond} = n_{\rm cond}^{(-)}$,
and this particle-number density plays the role of the order parameter.

In what follows we will investigate the phase structure of the
particle-antiparticle system with respect to the canonical variable $n_I$.
As a first step it is reasonable to obtain the dependencies of the densities
$n^{(-)}$, $n^{(+)}$ and mean field $U$ with respect to $n_I$ when we fix $T$.
Each isotherm crosses two different phases, so for each specific phase we must
solve the corresponding system of equations:
1) when both components of the pion gas are in the thermal (kinetic) phase,
it is necessary to solve Eqs.~(\ref{eq:tot-n}), (\ref{eq:ni});
2) when the $\pi^-$ subsystem of the pion gas has a condensate contribution,
equations (\ref{eq:tot-n2a}), (\ref{eq:ni2a}) are solved.
For the selected isotherm $T$, the point $n_I = n_{\rm Ic}$ divides the
$n_I$-axis into two parts.
When $n_I \le n_{\rm Ic}$, $\pi^-$- and $\pi^+$-mesons are in the thermal phase.
While for $n_I > n_{\rm Ic}$, the $\pi^-$-mesons have condensate contribution,
but the $\pi^+$-mesons are still completely in the thermal phase.

The dependence of the mean field on $n_I$ for three values of temperature,
$T = 40,\,80,\,100$~MeV, is shown in Fig.~\ref{fig:n-vs-ni-k01} on left panel.
It is seen that the difference in the curves associated with different temperatures
is very weak after the minimum of the function $U\big(n(T,n_I)\big)$.
In the point $n_I = n_{\rm I0}$, where $U(n_{\rm I0}) = 0$, the mean field
changes its sign  and becomes completely repulsive, at $\kappa = 0.1$
we obtained $n_{\rm I0} \approx 0.14$~fm$^{-3}$.

The results of calculation of the particle-number densities $n^{(-)}$,\, $n^{(+)}$
and $n \equiv n_{\rm tot} = n^{(-)} + n^{(+)}$ as functions of isospin density $n_I$
at $T = 101.7$~MeV, $\kappa = 0.1$
are depicted in Fig.~\ref{fig:n-vs-ni-k01} on right panel.
It is seen that the density of $\pi^+$ mesons decreases for small $n_I$
with increase of $n_I$ and then becomes approximately constant.
In fact, this behavior is quite understood.
Indeed, in accordance with selfconsistent Eqs.~(\ref{eq:tot-n}), (\ref{eq:ni})
or Eqs.~(\ref{eq:tot-n2a}), (\ref{eq:ni2a}) (it does not matter what pair of
equations we take) one has $n^{(+)} = (n - n_I)/2$.
But as we see the raise of the total particle density $n$ at the beginning is
much lower than $n_I$.
That is why, with increase of $n_I$ the particle density of $\pi^+$ mesons goes
down at the beginning and then becomes approximately constant.

\begin{figure}[h]
\centering
\includegraphics[width=0.49\textwidth]{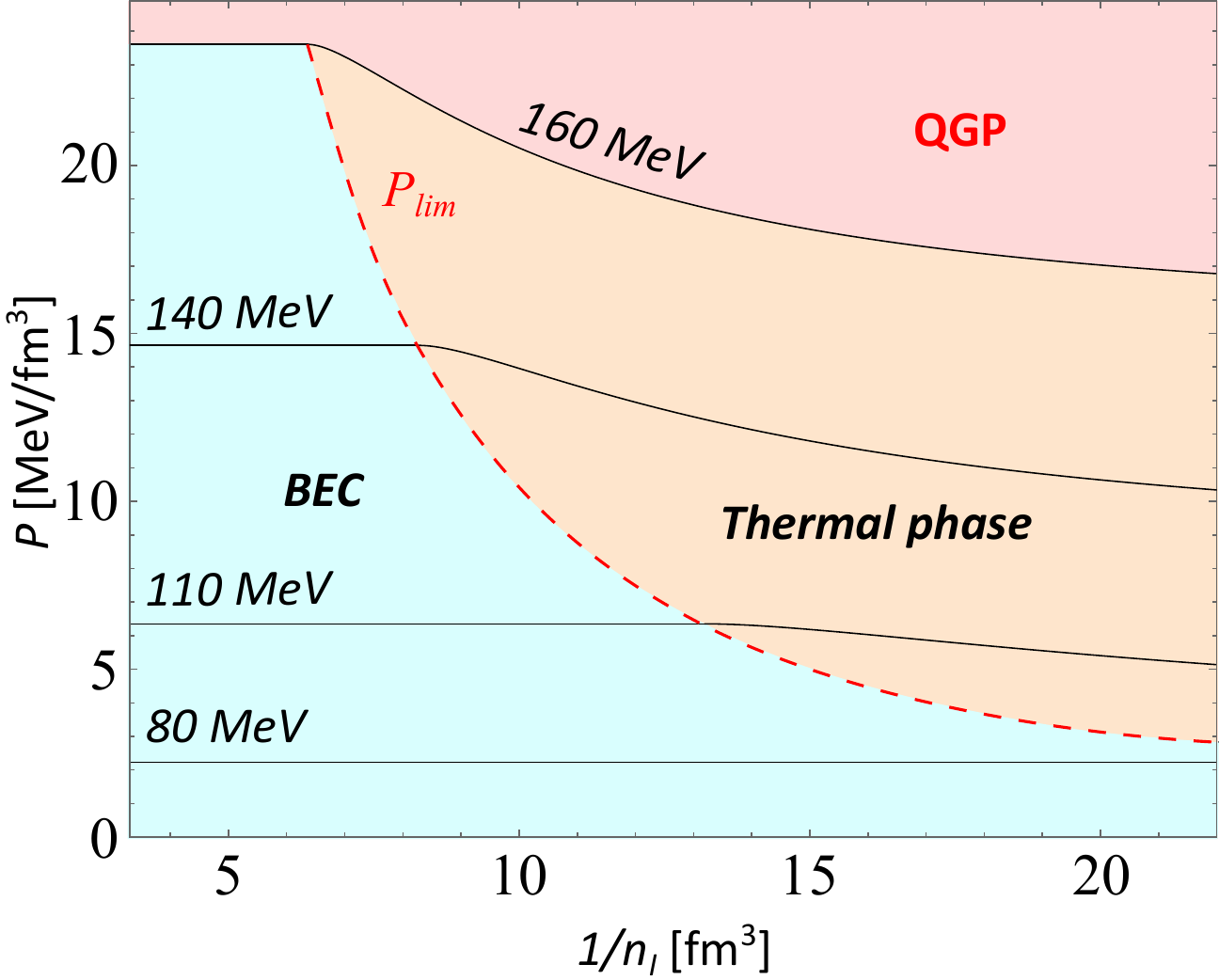}
\includegraphics[width=0.49\textwidth]{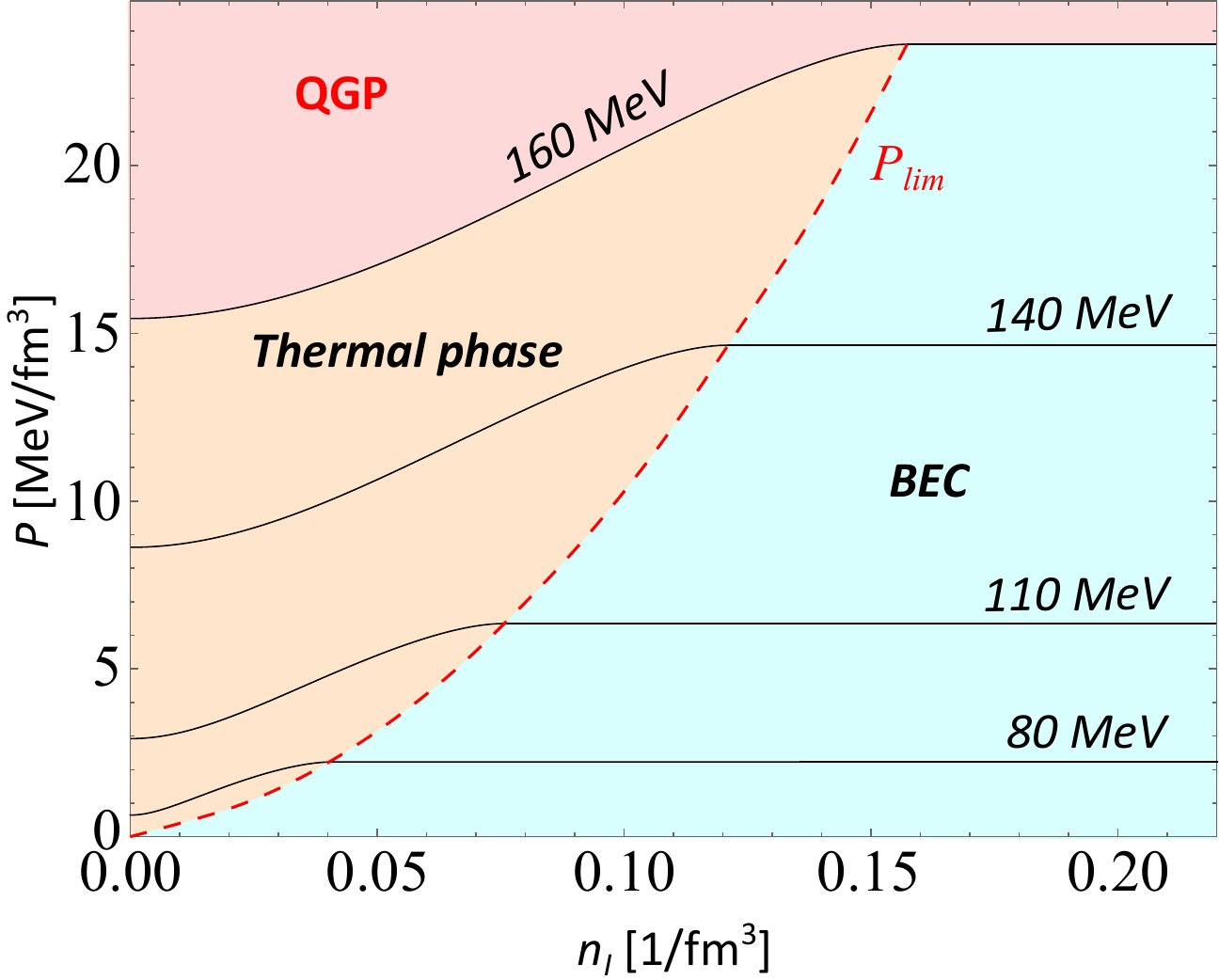}
\vspace{-1mm}
\caption{ Phase diagrams in an ideal $\pi^-$-$\pi^+$ gas.
Dependence of pressure on the inverse isospin density $v = 1/n_I$ (left panel)
and against the isospin density $n_I$ (right panel).
The $T = 160$~MeV isotherm is the approximate beginning of the QGP phase.
The red dashed curve $p_{\rm lim}$ indicates the pressure of a single-particle
ideal gas at $\mu = m$ and separates thermal phase from the BEC phase.
}
\label{fig:phase-diagram-ideal}
\vspace{10mm}
\includegraphics[width=0.49\textwidth]{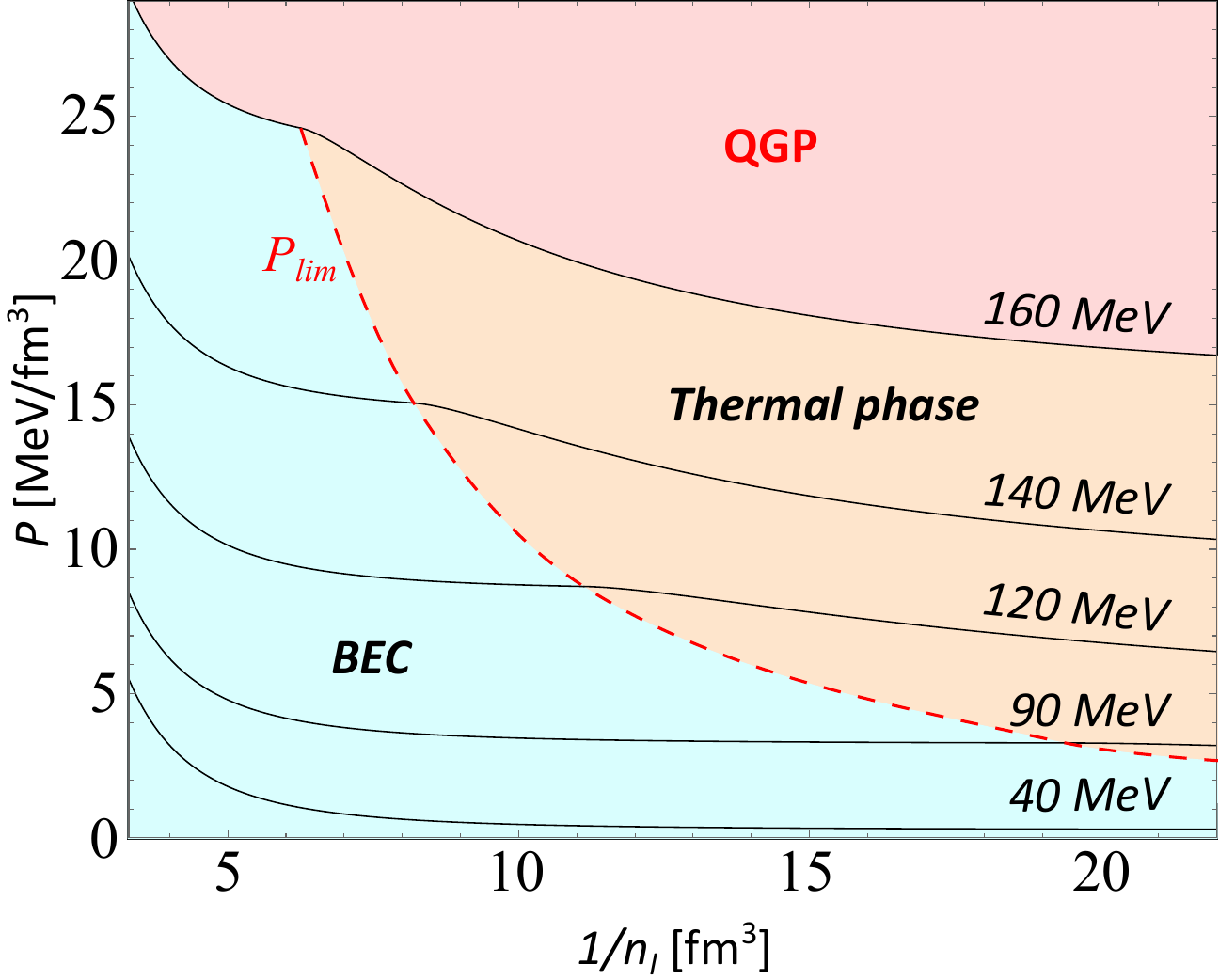}
\includegraphics[width=0.49\textwidth]{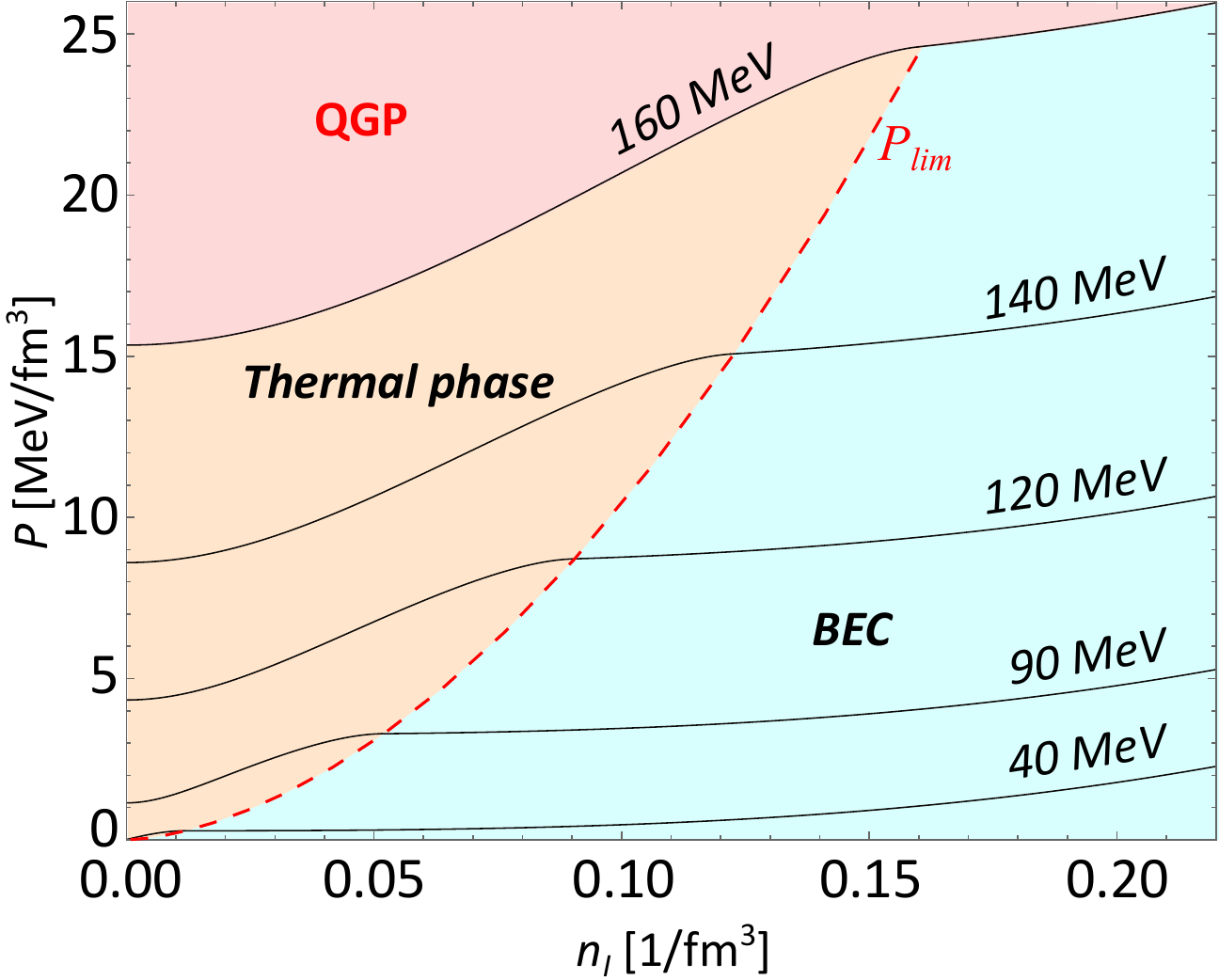}
\vspace{-1mm}
\caption{ Phase diagrams of the interacting $\pi^-$-$\pi^+$ system with
repulsion between particles, i.e., the attraction parameter $\kappa = 0$.
Dependence of pressure on the inverse isospin density $v = 1/n_I$ (left panel)
and on the isospin density $n_I$ (right panel).
The $T = 160$~MeV isotherm is the approximate beginning of the QGP phase.
The red dashed curve $p_{\rm lim}$ indicates the pressure on the critical curve
$n_{\rm lim}$, see eq.~(\ref{eq:nlim-id}).
}
\label{fig:phase-diagram-repulsion}
\end{figure}

\section{The phase diagram of the boson particle-antiparticle
\newline
system }
\label{sec:thrmod-properties}

In order to analyze the phase structure of the particle-antiparticle system,
we look for the dependence of the pressure $p(T,n_I)$ on the isospin density
$n_I$ at a fixed temperature $T$, that is, we study the behavior of the pressure
on the isotherm.
At the beginning, as a first step we test an ideal gas of $\pi^-$-$\pi^+$ mesons
as a reference point.
The phase structure of the system in this case is depicted in the two
panels in Fig.~\ref{fig:phase-diagram-ideal} for the dependencies
$p = p(v_I)$ and $p = p(n_I)$ , where $v_I = 1/n_I$.
In this figure the isotherm $T_{\rm qgp} = 160$~MeV separates the states of quark-gluon
plasma (QGP) and the blue shaded area marked BEC represents the states of the
Bose-Einstein condensate.
We see that there is no liquid-gas phase transition and the pressure of an
ideal gas naturally increases with increasing $n_I$ in the thermal phase,
but becomes constant in the condensed phase.
This effect arises because in a multi-particle system without interaction,
pressure exists only due to the kinetic movement of thermal particles with
nonzero momentum, $\bs k \ne 0$.
When we increase the particle density from zero and go along a specific isotherm
$T$ we come to the point $n_{\rm cd}$ on the critical curve
\footnote{The values $T_{\rm cd}$ and $n_{\rm cd}$ corresponds to the same point
on the critical curve $n_{\rm lim}(T)$.
The notation $T_{\rm c}$ is reserved for the critical isotherm in the description
of the liquid-gas phase transition.}.
In this point we rich the maximum density of the thermal particles.
Further increasing of the particle density is due only to increase of the
density of condensate particles.
At the same time, in the condensate phase, with an increase in $n_I$, the total
density of particles in the system increases only due to an increase in the
density of condensed particles with $\bs k = 0$, which does not contribute to
the pressure.
%
\begin{figure}
\centering
\includegraphics[width=0.49\textwidth]{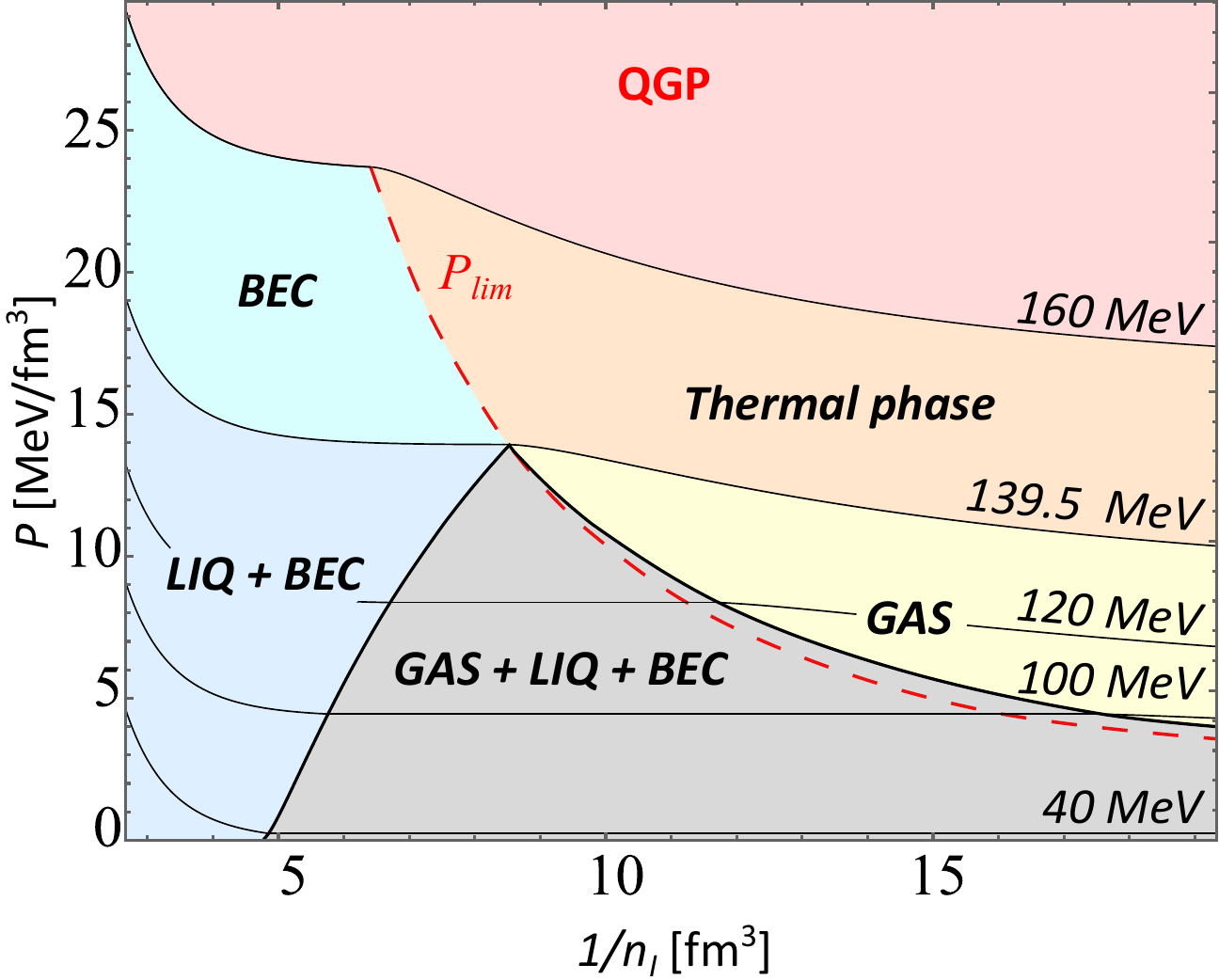}
\includegraphics[width=0.49\textwidth]{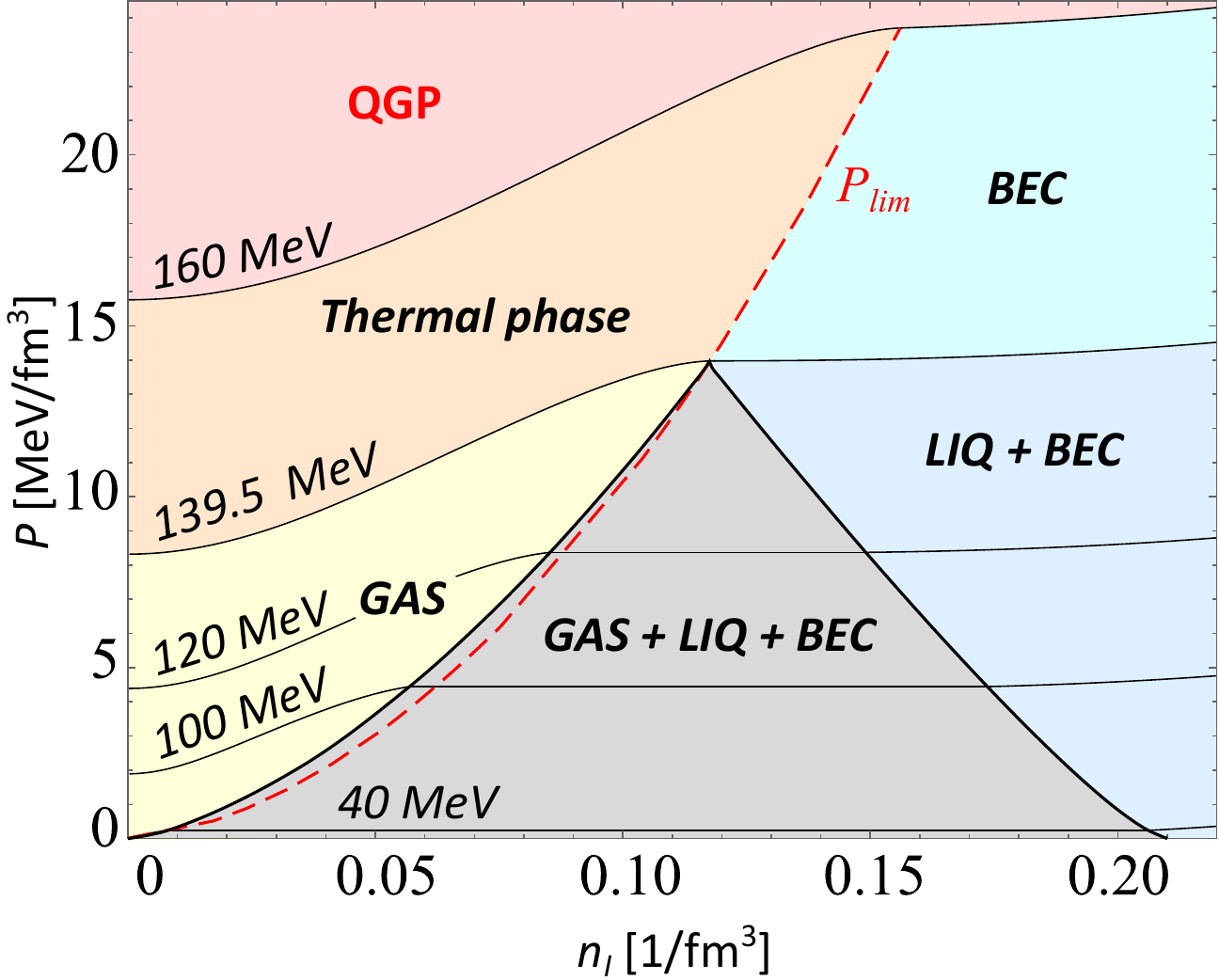}
\caption{ Phase diagrams: dependence of pressure on the inverse isospin density
$v = 1/n_I$ (left panel) and on the isospin density $n_I$ (right panel) in
$\pi^-$-$\pi^+$ interacting system at $\kappa = 0.2$.
The isotherm $T_{\rm qgp} = 160$~MeV  separates the QGP phase.
The red dashed curve $p_{\rm lim}$ indicates the pressure of an ideal gas at
$\mu = m$ and separates the thermal phase from the BEC phase.
The grey dashed ``triangle''\, represents the mixed phase of the gas and liquid,
which is almost in the $\pi^-$-meson condensate.
}
\label{fig:phase-diagram-k02}
\end{figure}

On the next step the system with only repulsion between particles,
i.e. $\kappa = 0$, was examined.
The phase structure of the system in this case is depicted on two panels
in Fig.~\ref{fig:phase-diagram-repulsion}.
In every panel we see three different phases: 1) ``Thermal phase'' - particles
and antiparticles are both in thermal states; 2) ``BEC'' - the subsystem of
particles, i.e. $\pi^-$ mesons, has the Bose-Einstein condensate contribution
and the subsystem of antiparticles, $\pi^+$ mesons, is in the thermal phase;
3) ``QGP'' - the phase where the quark-gluon plasma occurs, this phase is
separated by the isotherm $T = T_{\rm qgp} = 160$~MeV (we assume a melting
of all pion states at temperatures $T > T_{\rm qgp}$).
The line $p_{\rm lim}$ is the pressure of $\pi^-$ mesons on the critical curve
$n_{\rm lim}$ (see eq.~(\ref{eq:nlim-id})).
As can be seen in the region of ``condensate'', the behavior of isotherms
in the system of interacting bosons differs from isotherms in an ideal gas:
the pressure increases with an increase in isospin density.
This effect is due to the presence of positive excess pressure $P_{\rm ex}(n)$
as an additional contribution along with the kinetic pressure in the system.

%
\begin{figure}
\centering
\includegraphics[width=0.49\textwidth]{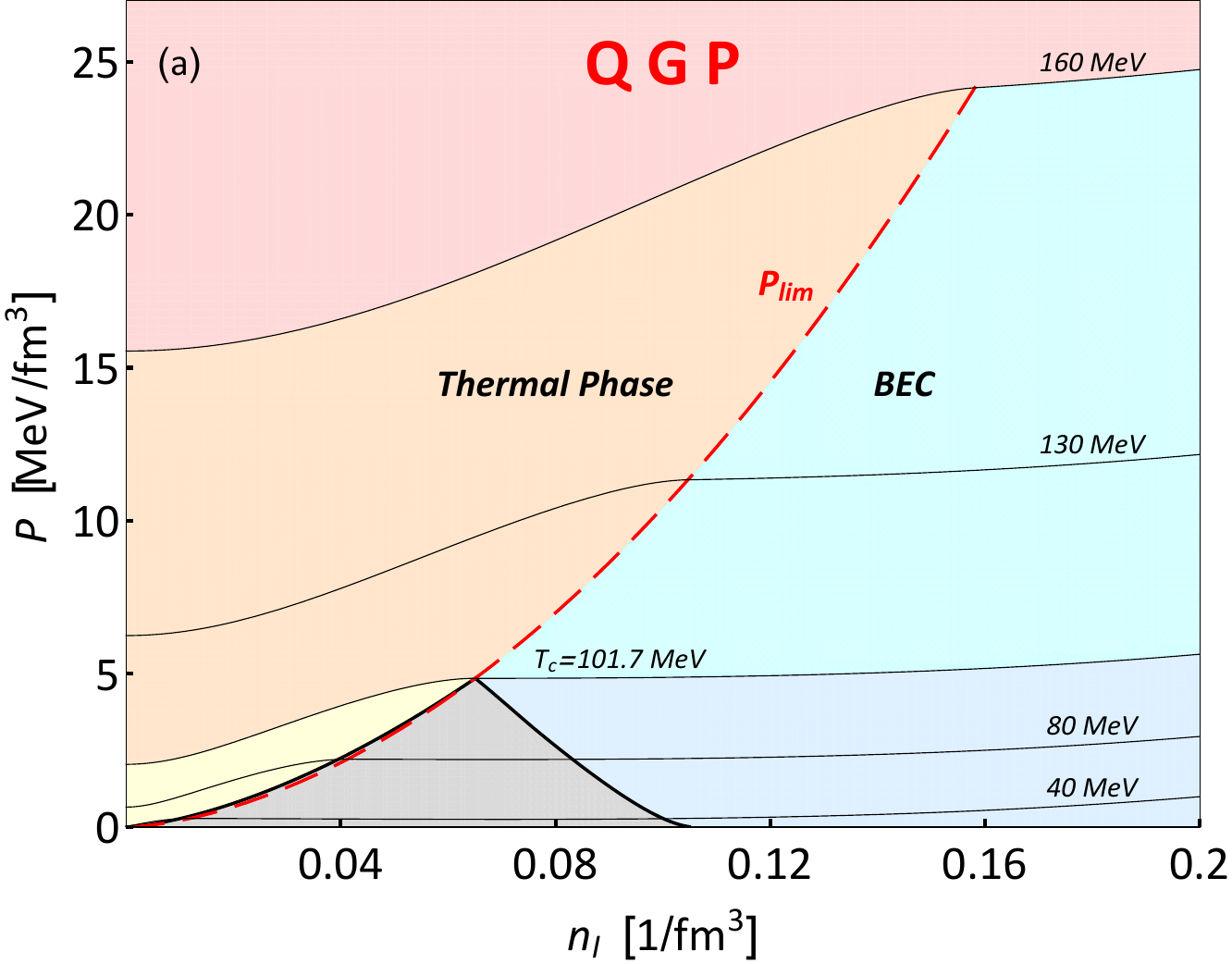}
\includegraphics[width=0.49\textwidth]{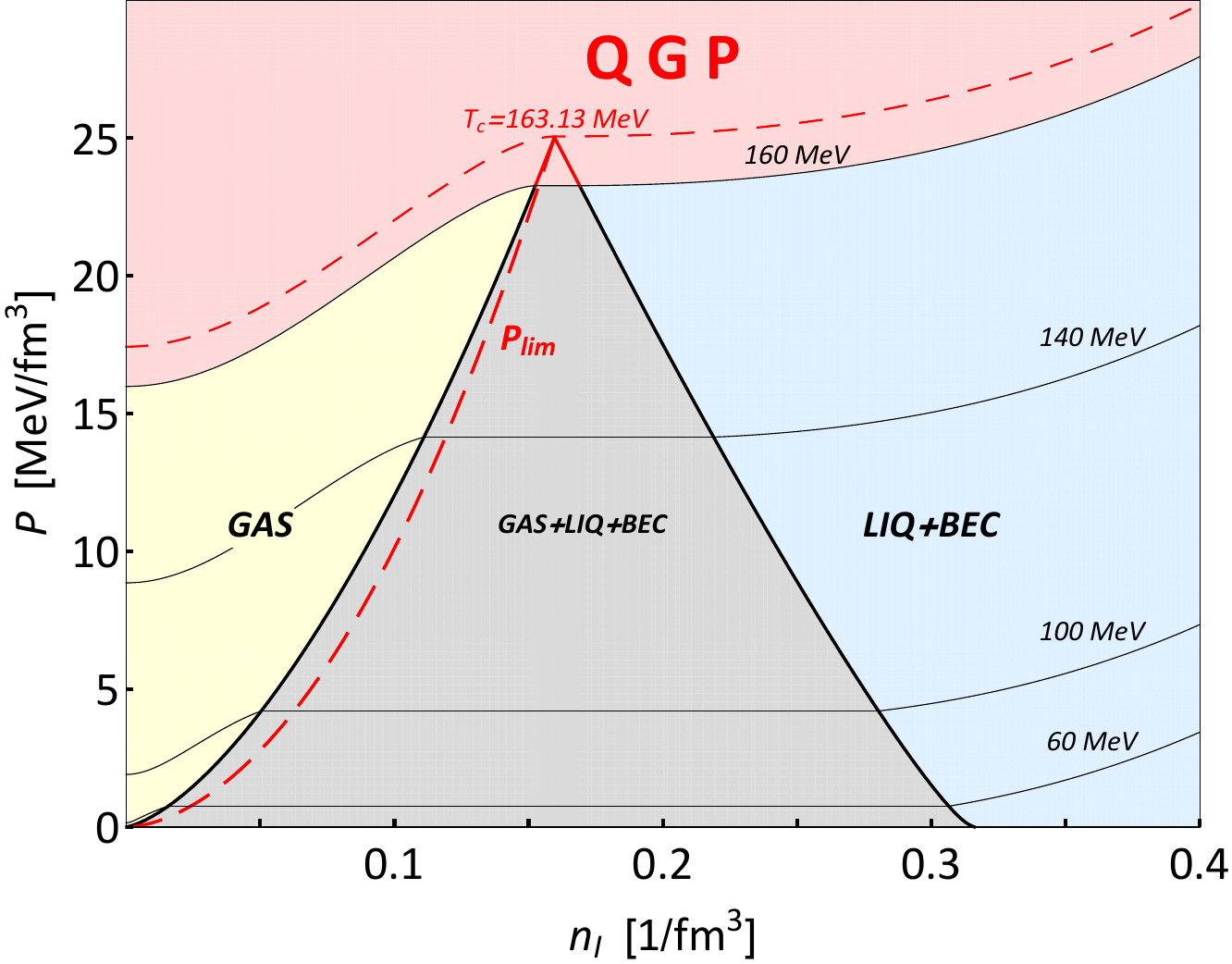}
\caption{ Phase diagrams: dependence of pressure on the isospin density in
the interacting $\pi^-$-$\pi^+$ system at $\kappa = 0.1$ (left panel) and
$\kappa = 0.3$ (right panel).
{\it Left panel:}
A liquid-gas phase transition occurs below the critical isotherm
$T_{\rm c} = 101.7$~MeV.
The isotherm $T = 160$~MeV  is the boundary of the QGP phase.
The red dashed curve $p_{\rm lim}$ separates the thermal phase from
the BEC phase.
{\it Right panel:}
The virtual critical isotherm $T_{\rm c} = 163$~MeV, calculated at
$\kappa = 0.3$, lies in the QGP phase, which is bounded by the isotherm
$T = 160$~MeV.
}
\label{fig:phase-diagram-k01}
\end{figure}

\subsection{The liquid-gas phase transition in the interacting particle-antiparticle
system of bosons}
\label{sec:dicussion-quantum-bosons}

In this section, we continue the discussion of an interacting two-component
bosonic system that includes both repulsions and attractions between particles.
If there is an attraction between particles, then the isotherms for temperatures
from the interval $T < T_{\rm c}$ show a wave or ``sinusoidal'' behavior in the
finite interval of $n_I$ ($T_{\rm c}$ determines the critical isotherm).
According to the standard thermodynamic approach, this specific isotherm behavior
can be considered as a liquid-gas phase transition.
To solve the problem, we apply Maxwell's generalized rules (see the Appendix),
which, unlike the textbook presentation of Maxwell's design, deal with the
isospin (charge) density rather than the total particle density.
As it follows from the generalized Maxwell's rules the pressure associated with
the isotherms, which cross the mixed liquid-gas phase, has a constant value,
as well as the chemical potential.
As a result we obtain the binodal, which determines the region of the liquid-gas
phase transition.
The resulting phase diagram is shown in Fig.~\ref{fig:phase-diagram-k02}.
The mixed liquid-gas phase (shaded gray area) is almost entirely in
the condensate phase (denoted as GAS+LIQ+BEC).
Recall that the condensate density in a two-component pion system
is created only by $\pi^-$ mesons, i.e. $n_{\rm cond} = n_{\rm cond}^{(-)}$.
This means that a certain part of $\pi^-$ mesons consists of particles with
$\bs k = 0$.
At the same time, thermal $\pi^-$ mesons together with $\pi^+$ mesons create
a kinetic mixture of gas and liquid.
Similar phase diagrams were calculated in \cite{satarov-2017} for alpha matter
and in \cite{anchishkin-2021} for pion matter.

If quantum statistics are taken into account when considering the liquid-gas
phase transition, then due to the appearance of condensate, the situation
becomes different than when describing the same model within the framework
of Boltzmann statistics.
It turns out that the presence of condensate strongly affects the position of
the local maximum of pressure.
Indeed, this maximum is localized now on the curve $p_{\rm lim}(n_I)$, which
represents the pressure in the system that is determined by the states belonged
to the critical curve $n_{\rm lim}(T)$ of the $\pi^-$-meson subsystem.
Remind, only this subsystem of mesons develops the Bose-condensate in case of the
weak attraction with $n^{(-)}\big|_{\rm crit. curve} = n_{\rm lim}(T)$.
The states $(T,n_{\rm lim}(T))$ on the critical curve corresponds to the
total density of particles $n = 2n_{\rm lim}(T) - n_I$.
For the temperature points on the critical curve $n_{\rm lim}(T)$ the
correspondence between temperature and the isospin density is determined
by equation $n_I = n^{(-)}(T) - n^{(+)}(T)$ that reads
\begin{equation}
n_I = n_{\rm lim}(T) - \int \frac{d^3k}{(2\pi )^3} \,f_{_{\rm BE}}\big( E(k,n),-\mu_I
\big)\Big|_{\mu_I = U(n) + m} \,,
\label{eq:nneg-ST2-3}
\end{equation}
where $U(n) = U\!\left( 2n_{\rm lim} - n_I \right)$ and $E(k,n) = \omega_k + U(n)$.
Solving this equation one gets the dependence $T = T(n_I)$, where $T$ is the point
on the critical curve $n_{\rm lim}(T)$ (see the red dashed line in
Fig.~\ref{fig:article-n-vs-T}).
The total pressure at these states reads
\begin{equation}
p_{\rm lim}(n_I) \,=\, p_{\rm kin}\big(n_I\big) \,+\, P_{\rm ex}(n)\,,
\label{eq:ptot-crit-curve2}
\end{equation}
where the kinetic pressure at the states $(T,n_{\rm lim}(T))$,
where $T= T(n_I)$, looks like
\begin{equation}
p_{\rm kin}(n_I) \,=\, \frac{1}{3} \int \frac{d^3k}{(2\pi)^3}\,\frac{{\bf k}^2}{\omega_k}
\big[ f_{_{\rm BE}}\big(\omega_k, m\big) \,
+\, f_{_{\rm BE}}\big(E(k,n),-\mu_I\big)\big]_{T= T(n_I)}
\label{eq:p-kin-crit-curve}
\end{equation}
with $E(k,n) = \omega_k +U(n)$, $\mu_I = U(n) + m$ and
$n = 2n_{\rm lim}\big(T(n_I)\big) - n_I$.
The curve $p_{\rm lim}(n_I)$ separates the pressure that correspond to the
condensate states (shaded area marked as BEC) from the pressure that correspond
to the thermal states of the boson system,
see Figs.~\ref{fig:phase-diagram-repulsion}-\ref{fig:phase-diagram-k01}.

Next, we are going to prove two features that are inherent to the behavior of
the pressure in the condensate phase.
\\
The first feature: The kinetic pressure along each isotherm in the
condensate phase is approximately constant
\begin{equation}
p_{\rm kin}(T,n_I)\big|_{T = {\rm const}}  \, \approx \, {\rm const} \,.
\label{eq:pkin}
\end{equation}
a) This will be an exact equality in the absence of interaction between particles
when $p \,=\, p_{\rm kin} + P_{\rm ex}$ reduces to $p = p_{\rm kin}(T,n_I)$.
The pressure of the two-component ideal gas is depicted
in Fig.~\ref{fig:phase-diagram-ideal}.
We see the constant pressure in the condensate phase (shaded blue area).
The effect, which is indicated in Eq.~(\ref{eq:pkin}),
arises because the increase of the variable $n_I$ in the condensate
phase occurs only due to the increase of condensed particles, whereas the
density of thermal particles remains constant along isotherm.
But, the increase of the number of particles in the system with zero momentum $\bs k = 0$
do not contribute to the kinetic pressure.
Here we discussed the pressure of $\pi^-$ mesons, which develop the condensate states.
The partial pressure of $\pi^+$ mesons on the same isotherm $T$ is
calculated for $\mu = m$ and it is created only by thermal particles, which
density is also constant in the condensate phase.
By this we prove the rigorous validity of Eq.~(\ref{eq:pkin}) in the ideal
$\pi^-$ - $\pi^+$ meson system.

\noindent
b) In the system with interaction equality (\ref{eq:pkin}) is approximately valid.
As it is seen from Eq.~(\ref{eq:p-kin-crit-curve}) the first contribution of
pressure, i.e., the kinetic partial pressure of $\pi^-$ mesons, which contributes
$98\, \%$ to $p_{\rm kin}$, is still constant when we increase $n_I$ because
$U(n) - \mu_I = - m$ in the condensate phase.
The kinetic partial pressure of $\pi^+$ mesons, which are in the thermal phase,
is suppressed because the distribution function looks like
$f_{_{\rm BE}} = 1/\{\exp{[(\omega_k + 2U(n) + m)/T]} - 1\}$ and a contribution
of $\pi^+$ mesons to kinetic pressure is not more than $2\, \%$.
Hence, we can adopt that in the condensate phase the kinetic pressure in
$\pi^-$ - $\pi^+$ meson system is constant with a good accuracy
and Eq.~(\ref{eq:pkin}) is approximately valid.

\begin{figure}
\centering
\includegraphics[width=0.32\textwidth]{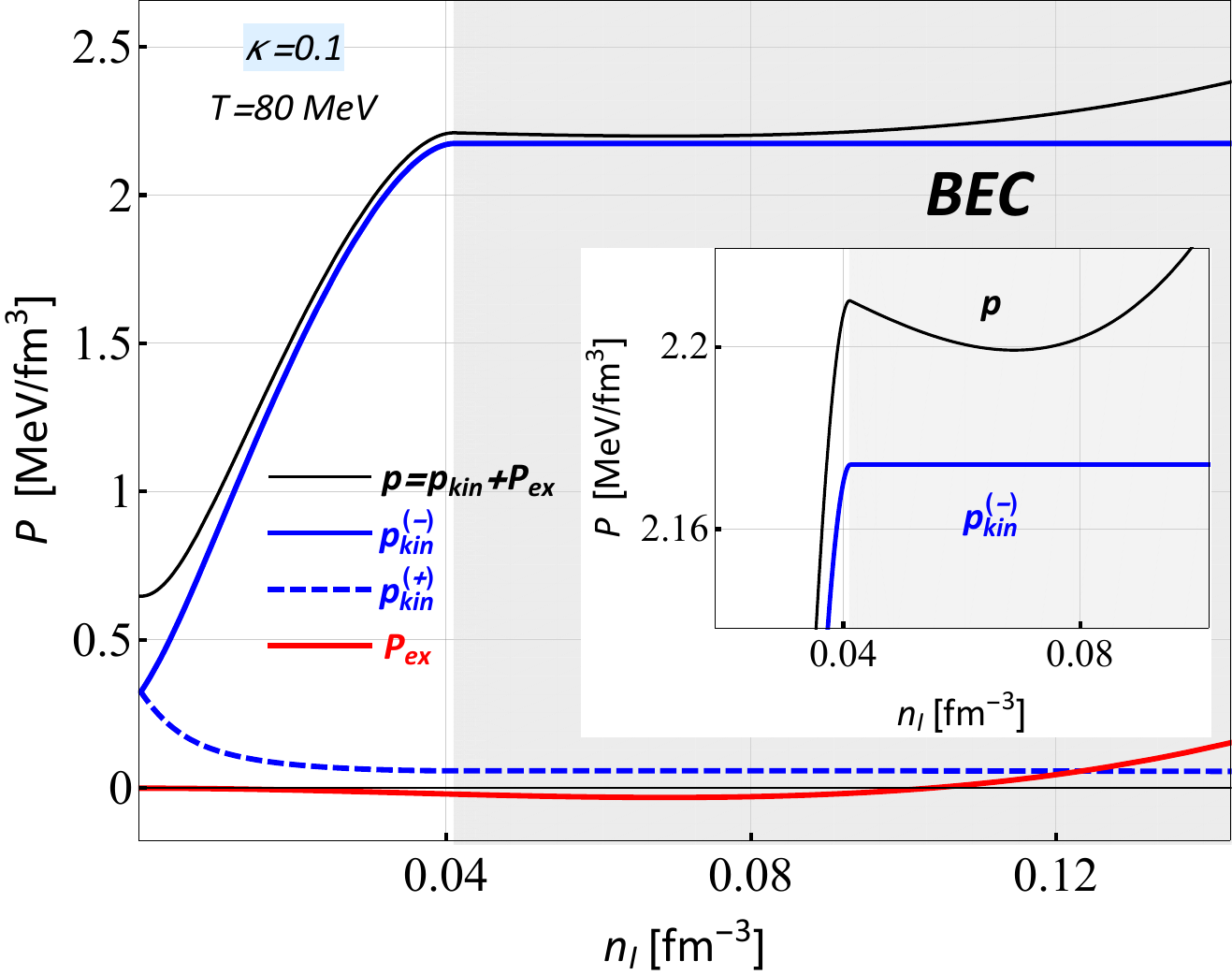}
\includegraphics[width=0.32\textwidth]{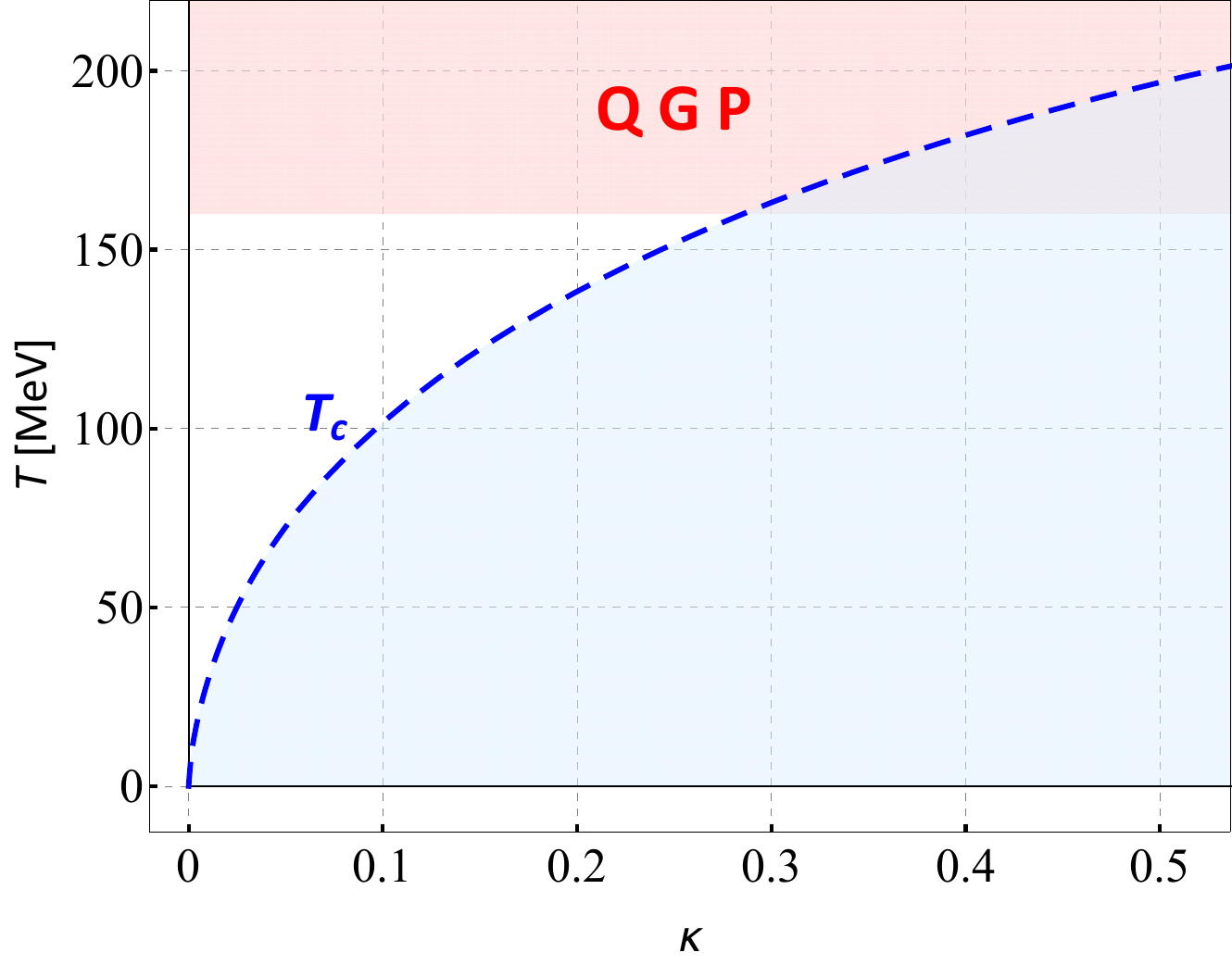}
\includegraphics[width=0.32\textwidth]{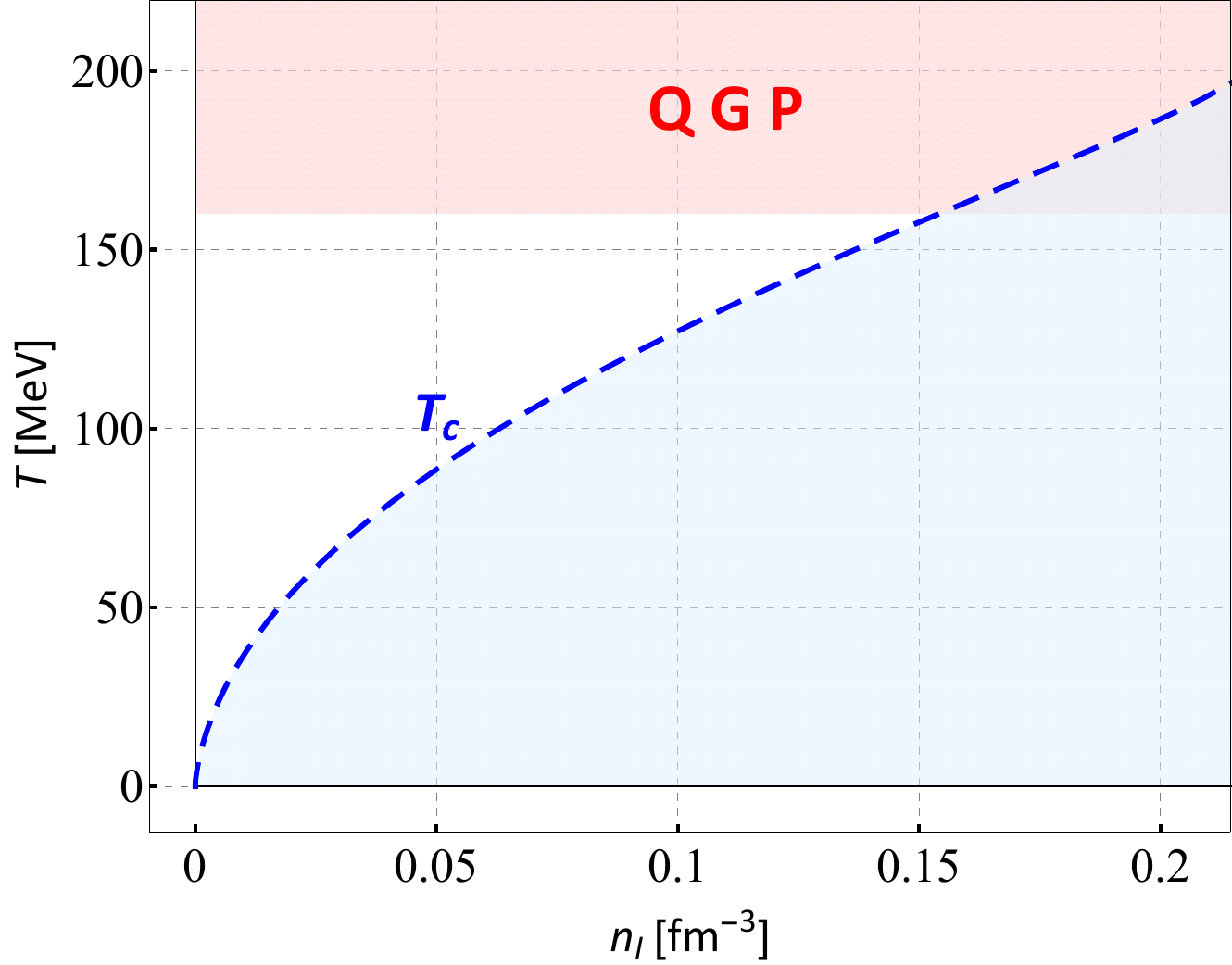}
\caption{
{\it Left panel:}
Dependence of partial pressure contributions on the isospin density in the
$\pi^-$-$\pi^+$ interacting system at $T = 80$~MeV and $\kappa = 0.1$.
The small window shows the local pressure maximum at the edge of the condensate
region and the local pressure minimum, which is due to the presence of the
excess pressure $P_{\rm ex}$.
The shaded area indicates Bose-Einstein condensate states (BEC).
{\it Center panel:}
Dependence of the critical temperature $T_{\rm c}$ of the liquid-gas phase
transition on the attraction parameter $\kappa$.
{\it Right panel:}
Dependence of the critical temperature $T_{\rm c}$ of the liquid-gas phase
transition on the isospin density $n_{\rm I}$.
}
\label{fig:tc-vs-kappa}
\end{figure}

The second feature:
The pressure has the following structure, $p = p_{\rm kin} + P_{\rm ex}$.
It is obvious that the kinetic pressure is always positive, $p_{\rm kin} > 0$,
and the excess pressure is not, its sign depends primarily on the density $n_I$.
As can be seen in Fig.~\ref{fig:n-vs-ni-k01} on the left panel, the excess
pressure $P_{\rm ex}$ is negative ($U(n)$ and $P_{\rm ex}$ have the same sign).
Due to this, with increasing of $n_I$ the pressure on each isotherm $T < T_{\rm c}$
begins to go down in the condensate phase after crossing the line $p_{\rm lim}(n_I)$.
This decreasing of the pressure is going on till the point of the local minimum,
as it is shown in Fig.~\ref{fig:tc-vs-kappa} on the left panel.
In this figure we present the results of calculation of the kinetic pressure
$p_{\rm kin}^{(-)}$ of the $\pi^{(-)}$ mesons, $p_{\rm kin}^{(+)}$ of the
$\pi^{(+)}$ mesons and the excess pressure $P_{\rm ex}$.

The liquid-gas phase transition occurs when the pressure in
the system possesses firstly a local maximum and then a local minimum when the
isospin density $n_I$ increases.
Let us consider the structure of the pressure in the condensate region.
For the derivative of the total pressure after some algebra
(see Appendix \ref{sec:deriv-pressure-cond}) we get
\begin{equation}
\frac{\partial p(T,n_I)}{\partial  n_I} \,
=\, \left( 1 - \frac{2 n^{(+)}}{n}  \right) \, \frac{\partial
P_{\rm ex}(n)}{\partial  n} \, \frac{\partial n}{\partial  n_I} \, =\, 0 \,.
\label{eq:minimum-pex-2}
\end{equation}
Here we use $\partial p_{\rm kin}^{(-)}(T,n_I)/\partial  n_I = 0$,
For a positive value of the bracket $(1 - 2 n^{(+)}/n) > 0$, where
$n = 2 n^{(+)} + n_I$, Eq.~(\ref{eq:minimum-pex-2}) leads to
\begin{equation}
\frac{\partial P_{\rm ex}(n)}{\partial  n} \,=\, 0
\qquad  \rightarrow  \qquad
n^{\rm (min)} \,=\, \frac{A}{2B} \,=\, \kappa \, \sqrt{\frac{m}{B}} \,,
\label{eq:n-pmin}
\end{equation}
where we use the explicit form of $P_{\rm ex}$ given in Eq.~(\ref{eq:pex-skyrme})
and $A = 2 \kappa \sqrt{mB}$.
Therefore, the local minimum of the pressure is in the point $n^{\rm (min)}$.
Indeed, it is a minimum, because the sign of the second derivative in this
point is positive, $\frac{\partial^2 P_{\rm ex}(n)}{\partial  n^2} = A > 0$.
On the other hand, at the critical temperature $T_{\rm c}$ the local minimum of
the pressure, $p_{\rm min}$, coincide with the local maximum, $p_{\rm max}$.
Conventionally, we can write this condition as
\begin{equation}
p_{\rm max}(T_{\rm c}) \,=\, p_{\rm min}(T_{\rm c}) \,.
\label{eq:pmax-pmin}
\end{equation}
As it was shown, the local minimum of the pressure is determined by the total
density $n^{\rm (min)}$ and, as it follows from Eq.~(\ref{eq:n-pmin}),
this density does not depend on the temperature.
It can be concluded that the local minimum of all isotherms corresponds to
the value $n^{\rm (min)}$ of the total number density, and it defines the
right wing of the spinodal.
At the same time, it gives us the value of the total number density at the
critical point
\begin{equation}
n_{\rm c} \,=\, \frac{A}{2B} \,=\, \kappa \, \sqrt{\frac{m}{B}} \,.
\label{eq:nc}
\end{equation}
In this point, it is appropriate to consider the model under consideration
within the framework of Boltzmann statistics: $p = Tn - (1/2)n^2 + (2/3)n^3$,
where we use the excess pressure (\ref{eq:pex-skyrme}).
The particle density $n_{\rm c}^{\rm B}$ at the critical point can be
determined by equation: $[\partial^2 p(T,n)/\partial n^2]_T = 0$.
We get that $n_{\rm c}^{\rm B} = A/4B$.
Comparing this value with the critical total density $n_{\rm c}$ in a bosonic
two-component system, we can point out the interesting fact that these values
are related to each other exactly as $2:1$,
i.e. $n_{\rm c} = 2 n_{\rm c}^{\rm B}$.
This relation was also mentioned in Ref.~\cite{gorenstein-2022}.

Now we will consider the peculiarities of determining the local pressure maximum
in a two-component bosonic system during the liquid-gas phase transition.
In the case of condensate in the system, it is impossible to determine the local
maximum of the pressure using the same equation
$\partial p(T,n_I)/\partial n_I = 0$, since it only determines the local minimum.
The isotherm is not a smooth function on the edge of the condensate, to the left
of its point of local minimum, as it was in the standard van der Waals picture.
This is explained by the specific behavior of pressure in the condensate phase.
According to the textbook procedure, to find the maximum or minimum of a smooth
function in a region with edges (finite interval), it is necessary also to
consider the value of the function at the edges of the region.

As we argued above,
the local maximum is at the edge of the condensate region, since starting from
the point of intersection with the condensate edge $p_{\rm lim}$ the pressure
on the isotherm $T \le T_{\rm c}$ begins to decrease, since
$p_{\rm kin} \approx$~const~$> 0$, but $P_{\rm ex} < 0$.
That is, to the left of the local minimum, we get a local maximum of the
pressure
on the $T \le T_{\rm c}$ isotherm, see Fig.~\ref{fig:tc-vs-kappa}, left panel.
On the graphs of the phase diagrams, see Fig.~\ref{fig:phase-diagram-k01},
the boundary between the thermal phase and the condensate phase is marked as
the pressure $p_{\rm lim}(n_I)$, which corresponds to the states on the critical
curve $n_{\rm lim}(T)$ as was defined in Eq.~(\ref{eq:p-kin-crit-curve}).

Now we will discuss the algorithm for calculating the critical temperature
$T_{\rm c}$.
Let us look at eq.~(\ref{eq:nc}) where we are dealing with the total
particle density.
According to solutions of self-consistent equations in the thermal and
condensate phases for a system with a conserved isospin density,
the total number density is related to the densities of $\pi^-$ mesons
and $\pi^+$ mesons as
$n(T,n_I) = n^{(-)}(T,n_I) + n^{(+)}(T,n_I)$.
For states on the critical curve, where $n^{(-)}(T) = n_{\rm lim}(T)$,
the total number density looks like: $n(T) = n_{\rm lim}(T) + n^{(+)}(T)$,
where $T = T(n_I)$.
Then, the total number density, which determines the local maximum of the
pressure, can be conventionally written as
$n(T_{\rm c}) = n_{\rm lim}(T_{\rm c}) + n^{(+)}(T_{\rm c})$.
We are looking for a point where the local maximum of the pressure coincides
with the local minimum of the pressure, which is determined by the critical
density $n_{\rm c}$  defined in eq.~(\ref{eq:nc}).
Thus, we arrive at the equation, which reflects an equality of the particle
densities that determines the local maximum and the local minimum of the
pressure.
This equation defines the critical temperature $T_{\rm c}$ of the
liquid-gas phase transition:
\begin{equation}
n_{\rm lim}(T_{\rm c}) + n^{(+)}(T_{\rm c})\,=\, n_{\rm c}\,,
\label{eq:tc-on-kappa}
\end{equation}
where $n_{\rm c} = A/2B = \kappa \sqrt{m/B}$ and $n_{\rm lim}(T)$ is defined
in Eq.~(\ref{eq:nlim-id}).
Here, the mean field $U(n)$, which is present in the number density $n^{(+)}$,
is also determined at the total density $n_{\rm c}$
\begin{equation}
n^{(+)}(T_{\rm c}) \,=\, \int \frac{d^3k}{(2\pi)^3}\, \frac{\dis 1}
{\dis \exp{\left\{ \big[\omega_k + 2U\!\left(n_{\rm c}\right)
+ m\big]/T_{\rm c} \right\}} - 1} \,,
\label{eq:nplus-tc}
\end{equation}
where we use that $\mu_I = U(n) + m$.
The solution of eq.~(\ref{eq:tc-on-kappa}) determines the dependence of
the critical
temperature on the attraction parameter $\kappa$, $T_{\rm c}(\kappa)$,
which is shown in Fig.~\ref{fig:tc-vs-kappa} in the central panel, with
blue dashed line.
%
%

On the other hand one can solve eq.~(\ref{eq:tc-on-kappa}) with respect to
attraction parameter and obtain dependence $\kappa(T_{\rm c})$.
Next, we can take into account that according to solutions of self-consistent
equations for a system with a conserved isospin density, the total number density,
which determines the local maximum of the pressure,
can be written as $n(T_{\rm c}) = 2 n_{\rm lim}(T_{\rm c}) - n_I$.
Then, we rewrite eq.~(\ref{eq:tc-on-kappa}) in the form
\begin{equation}
2 n_{\rm lim}(T_{\rm c}) - n_I \,=\, n_{\rm c}\,,
\label{eq:tc-on-kappa-2}
\end{equation}
where we take $n_{\rm c} =  \kappa(T_{\rm c}) \sqrt{m/B}$.
The solution of eq.~(\ref{eq:tc-on-kappa-2}) determines the dependence of
the critical temperature on the isospin density $n_I$, $T_{\rm c}(n_I)$.
These dependence is plotted in Fig.~\ref{fig:tc-vs-kappa} in the right
panel, with blue dashed line.

\subsection{Discussion}
\label{sec:discussion-2}

We can compare the characteristics of the liquid-gas phase
transition in the classical and quantum interacting boson systems.
The interaction in both systems was studied in the framework of the mean-field
model with the mean field $U(n) = - An + Bn^2$, at the same values
of attractive coefficient $A$ and repulsive coefficient $B$.
As a result we obtain that the presence of the condensate, which is due to
the quantum Bose statistics, sufficiently increases the value of
the critical temperature of the liquid-gas phase transition.
Indeed, at the attraction parameter $\kappa = 0.2$ for the classical gas we get
the critical temperature $T_{\rm c} = 2.8$~MeV ($T_{\rm c} = A^2/8B$)
and for the quantum system we get $T_{\rm c} = 139$~MeV.
Let us discuss this phenomena.

It is necessary to point out that in the quantum bosonic system the increase
in $T_{\rm c}$ comparing to the classical system is caused primarily by the
peculiarities of the pressure behavior in the condensate phase.
An additional contribution is due to an increase of effective attraction
``produced'' by the boson quantum statistics.
Indeed, in a quantum system the Bose-statistics ``generates'' an effective
attraction between particles, whereas the Fermi-statistics ``generates''
an effective repulsion.
Hence, the Bose-statistics
leads to an effective increasing of the the attraction coefficient $A$ in
the mean field $U(n)$.
Indeed, the statistically induced interaction potential $V(r)$ for the first
order quantum correction reads
\begin{equation}
V(r) = - T \ln{\left[ 1 \mp e^{- 2\pi r/\Lambda^2} \right]}  \,,
\label{eq:eff-potential}
\end{equation}
where $r$ is the distance between particles, $\Lambda = \sqrt{2\pi/mT}$ is
the thermal wave length and the upper sign corresponds
to the Fermi statistics and the lower sign to the Bose statistics.
Evidently, this two-particle effective potential possesses the attractive
behavior in case of the Bose statistics and the repulsive behavior in case of
the Fermi statistics (the correction is valid when $r \gg \Lambda$).
Hence, the transfer from the classical gas to the quantum one is twofold.
First, the effective attraction between particles goes up, and second
the condensate that is presented now in the system determines the increase of
the critical temperature.
The latter was also shown in Refs.~\cite{anchishkin-2021,gorenstein-2022}.
We emphasize that a presence of the condensate that is due to the quantum
statistics and the interplay of attraction and repulsion between particles
make this ``jump up'' of the critical temperature.
The mane reason for this is that, the kinetic pressure in the condensate is
almost constant in the interval of the liquid-gas phase transition.

At the same time, we can name the opposite effect which is due to the Fermi
statistics.
As was shown in Ref.~\cite{vovchenko-anchishkin-gorenstein-2015}, the quantum
van der Waals equation that was applied for description of the nuclear matter
gives for the liquid-gas phase transition the critical temperature $T_{\rm c} = 19.7$~MeV.
This value of the critical temperature is close to the experimental estimates
given in Refs.~\cite{natowitz-2002,karnaukhov-2003}.
On the other hand, the classical van der Waals equation gives for the liquid-gas
phase transition the value $T_{\rm c} = 29$~MeV.
As we discussed in this section and as it follows from Eq.~(\ref{eq:eff-potential}),
the Fermi statistics effectively amplifies the repulsion between particles
what means an effective decreasing of the attraction between particles.
This quantum-statistical effect of the decreasing of the attraction in the system
can be explanation of the decreasing of the
critical temperature from the value $T_{\rm c} = 29$~MeV to the value
$T_{\rm c} = 19.7$~MeV when moving from the classical description of the liquid-gas
phase transition in the nuclear matter to the quantum-statistical one.

\section{Discussion and concluding remarks}
\label{sec:conclusions}

The article presents a thermodynamically consistent approach for description of
the Bose-Einstein and liquid-gas phase transitions in a dense selfinteracting
bosonic system at a conserved isospin (charge) density $n_I$.
As an example we considered the system of meson particles with $m = m_\pi$ and
zero spin, we name these bosonic particles as ``pions'' just conventionally.
This choice was made because the charged $\pi$-mesons are the lightest nuclear
particle and the lightest hadrons that couple to the isospin number.
For the same reason a ``temperature creation'' of the particle-antiparticle
pairs in the temperature interval $T \le 200$~MeV becomes a common problem
for the quantum-statistical methods.
Description of thermodynamic properties of the system was carried out
using the Canonical Ensemble formulation, where the chemical
potential $\mu_I$ is a thermodynamic quantity which depends on the
canonical variables $(T,n_I)$.
To obtain phase diagram, which reflects the liquid-gas phase transition, we
calculated the dependence of pressure with respect to the isospin
density for different isotherms and then, we modified the pressure dependence
in accordance with the generalized Maxwell rules (see Appendix A).

It should be noted that the electrical charge of the condensate is negative
in case if the total charge of the system is negative.
And vice versa, the electric charge of the condensate would be positive if
the total charge of the bosonic system is positive.
It was shown that at a fixed temperature the dependencies of the particle densities
$n^{(-)}(T,n_I)$ and $n_{tot}(T,n_I)$  with respect to $n_I$ are almost linear and
close to one another for $n_I > n_{Ic}$.
It happens since for every fixed $T$ the value of
particle-number density $n^{(-)}_{\rm th}$ of thermal (kinetic) $\pi^-$-mesons
does not change ~\cite{anch-2022-prc} and the value of $n^{(+)}$
is small and approximately constant (see Fig.~\ref{fig:n-vs-ni-k01}, right panel).
Because only $\pi^{-}$ mesons undergo the phase transition to the
Bose-Einstein condensate, the increase of the densities  $n^{(-)}$
and $n = n_{tot}$ for $n_I > n_{Ic}$ is almost due to an increase of the density
of condensate.

Phase diagrams were introduced on Fig.~\ref{fig:phase-diagram-repulsion}.
The scale parameter of the
model  $\kappa = A/(2\sqrt{mB})$, which is itself a combination of
the mean-field parameters $A$ and $B$ ($U(n)=-An+Bn^2$) and the
particle mass, determines the different possible phase scenarios
which occur in the particle-antiparticle boson system.
When the attraction coefficient $A = \kappa A_c$, where $A_c \equiv 2\sqrt{mB}$
is zero (that is, $\kappa = 0$), the system can only be in thermal and
condensate phases without a liquid-gas phase transition.

In the case of $\kappa > 0$, liquid-gas phase transition occurs in the system,
and a transition from the thermal phase to the condensate one is
possible both with liquid-gas phase transition if $T < T_{\rm c}$ ,
and without it when $T \ge T_{\rm c}$.
In other words there is a region in the phase diagram where the BEC and
the mixed liquid-gas phase exist simultaneously (grey area on the left panel
in Fig.~\ref{fig:phase-diagram-k01}).
A similar situation is described in \cite{satarov-2017}, where the Bose-Einstein
condensation and the liquid-gas phase transition in $\alpha$-matter
were investigated.
Area above isotherm $T = T_{\rm qgp} = 160$~MeV (QGP) is the phase where the
quark-gluon plasma occurs.
We assume this to be a limitation of our model since a melting of all pion
states at temperatures higher than $T_{\rm qgp}$.

The role of neutral pions is left beyond the scope of the present paper.
The present analysis can be improved by addressing these issue in more detail
and also by generalizing the calculation to nonzero contribution
to the mean field which depends on $n_I$.
Authors plan to consider these problems elsewhere.

\section*{Acknowledgements}
The work of D.~A. and D.~Zh. is supported by the Simons Foundation and by the
Program "The structure and dynamics of statistical and quantum-field
systems" of the Department of Physics and Astronomy of the NAS of Ukraine.
D.A. appreciates discussions with M.~Gorenstein, I.~Mishustin and H.~St{\"{o}}cker.
This work is supported also by Department of target training of Taras Shevchenko
Kyiv National University and the NAS of Ukraine, grant No. 6$\Phi$-2024.

\appendix

\section{Liquid-gas phase transition: the Maxwell rules
in the system with conservation of the charge}
\label{sec:maxwell-rule}

To determine the points $v = 1/n_I$  which correspond to the
Maxwell rule one needs to solve the system of two equations with respect to
the points $v_{x1} = 1/n_{I2}$ and $v_{x2} = 1/n_{I1}$
(note, we obtain the pressure as function of $n_I$, i.e. $p(n_I)$).
In case of the homogeneous system for isothermal process $T =$~const we have
\begin{equation}
dp \,=\, sdT + n_I d\mu_I  \qquad  \rightarrow  \qquad
dp \,=\, n_I d\mu_I\,.
\label{eq:hibbs-hugel-3a}
\end{equation}
Then (1):
\begin{equation}
d\mu_I = 0 \qquad  \rightarrow  \qquad
dp \,=\, 0   \quad  \rightarrow  \quad   \int_1^2 dp = 0
\quad  \rightarrow  \quad   p_1 \,=\, p_2 \,.
\label{eq:hibbs-hugel-3b}
\end{equation}
Then (2):
\begin{equation}
\frac{1}{n_I} \, dp \,=\, d\mu_I \,, \qquad
d\mu_I = 0 \qquad  \rightarrow  \qquad
 \int_1^2 \frac{1}{n_I} \, dp = 0 \,.
\label{eq:hibbs-hugel-3c}
\end{equation}
We write these two equations (\ref{eq:hibbs-hugel-3b}),
(\ref{eq:hibbs-hugel-3c}) as a set with respect to $v_1$ and $v_2$,
\begin{eqnarray}
\label{eq:vx-1}
p(1/v_1) &=& p(1/v_2) \,,
\\
\int_{v_1}^{v_2} dv \, p(1/v) &=& (v_2 - v_1)\,p(1/v_1) \,,
\label{eq:vx-2}
\end{eqnarray}
where last equation can be read as
\begin{equation}
\int_{v_1}^{v_2} dv \, \big[ p(1/v) - p(1/v_1) \big] = 0 \,.
\label{eq:single-eq-mr}
\end{equation}
Then, after we get solution one can determine the values $n_{I1} =
1/v_2$ and $n_{I2} = 1/v_1$.

\begin{figure}[h!]
\centering
\includegraphics[width=0.24\textwidth]{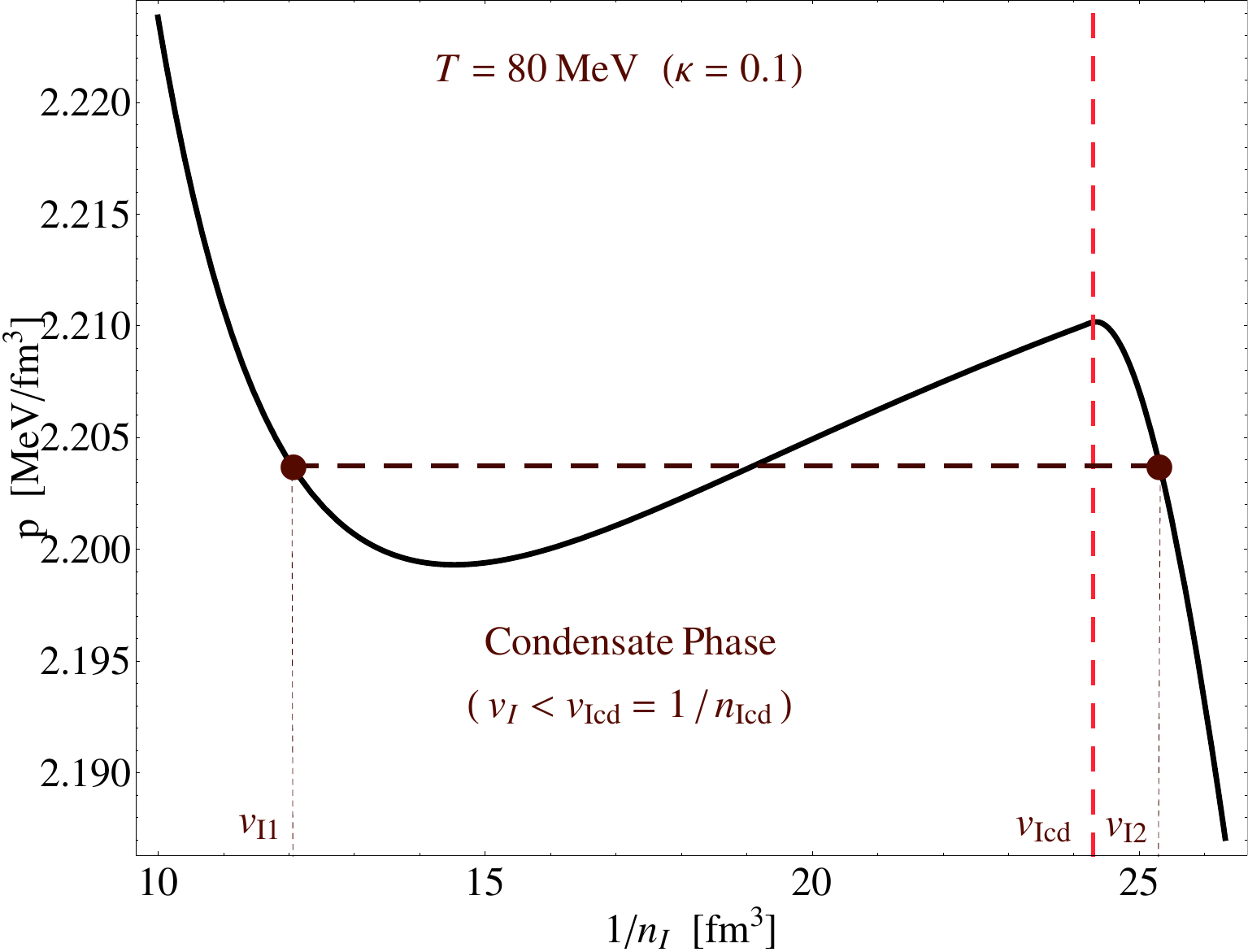}
\includegraphics[width=0.24\textwidth]{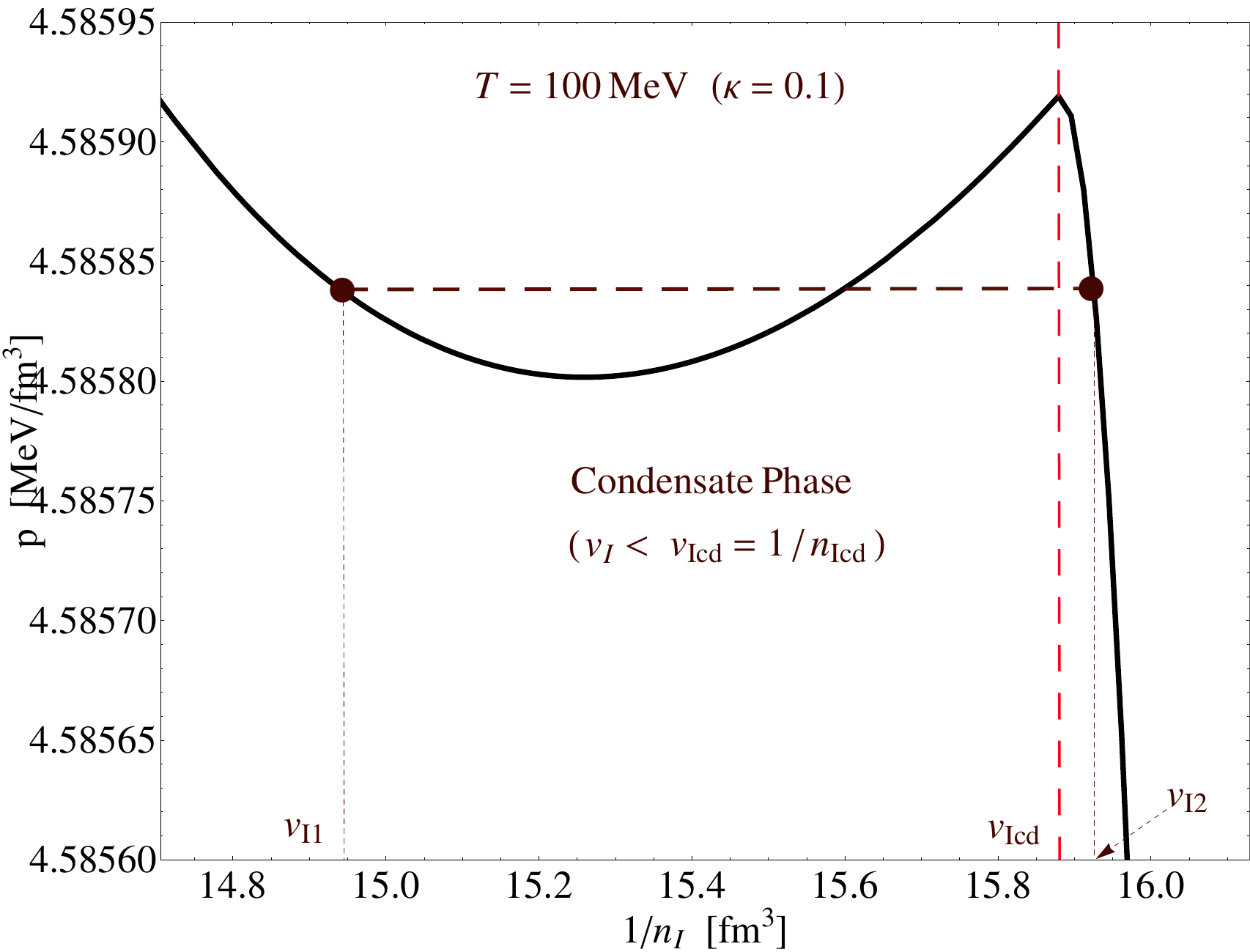}
\includegraphics[width=0.24\textwidth]{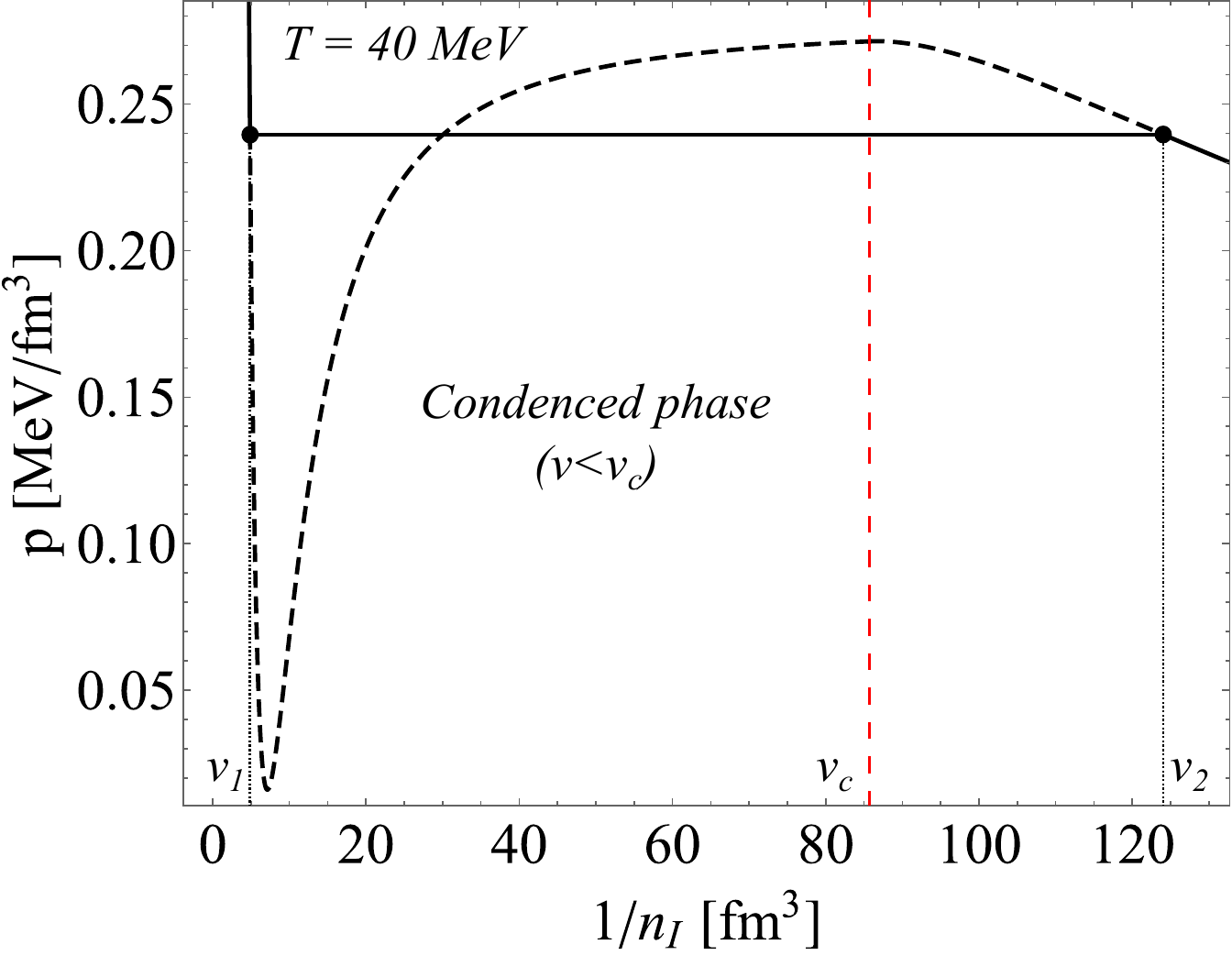}
\includegraphics[width=0.24\textwidth]{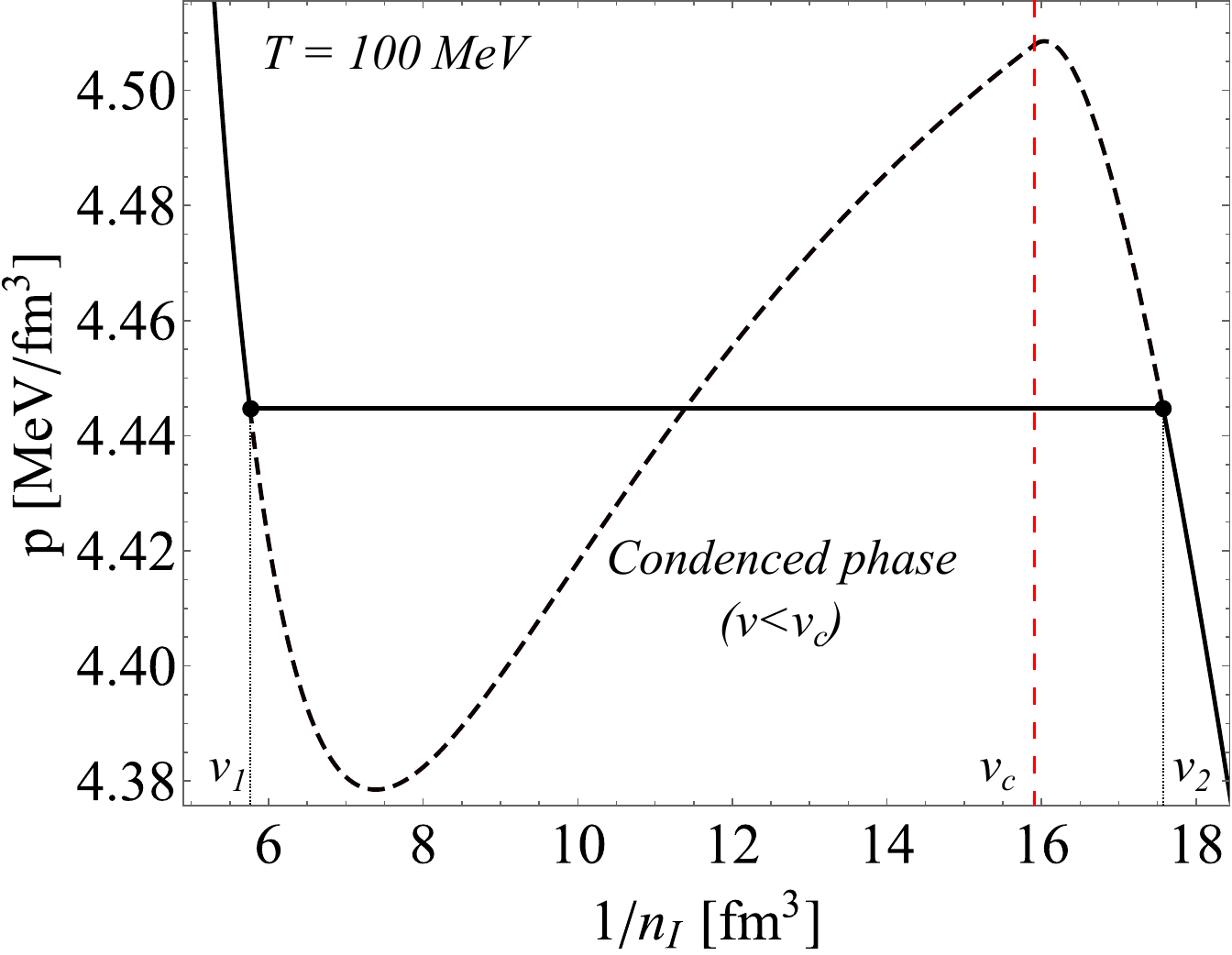}
\caption{
Dependence of pressure on the inverse isospin density $1/n_I$  in the interacting
$\pi^+$-$\pi^-$ pion system within the framework of the mean-field model:
in the two left panels at $\kappa = 0.1$ and in the two right panels at
$\kappa = 0.2$.
The liquid-gas phase transition is determined by Maxwell's generalized rules.
}
\label{fig:p-vs-invnI}
\end{figure}

\begin{figure}[h]
\centering
\includegraphics[width=0.24\textwidth]{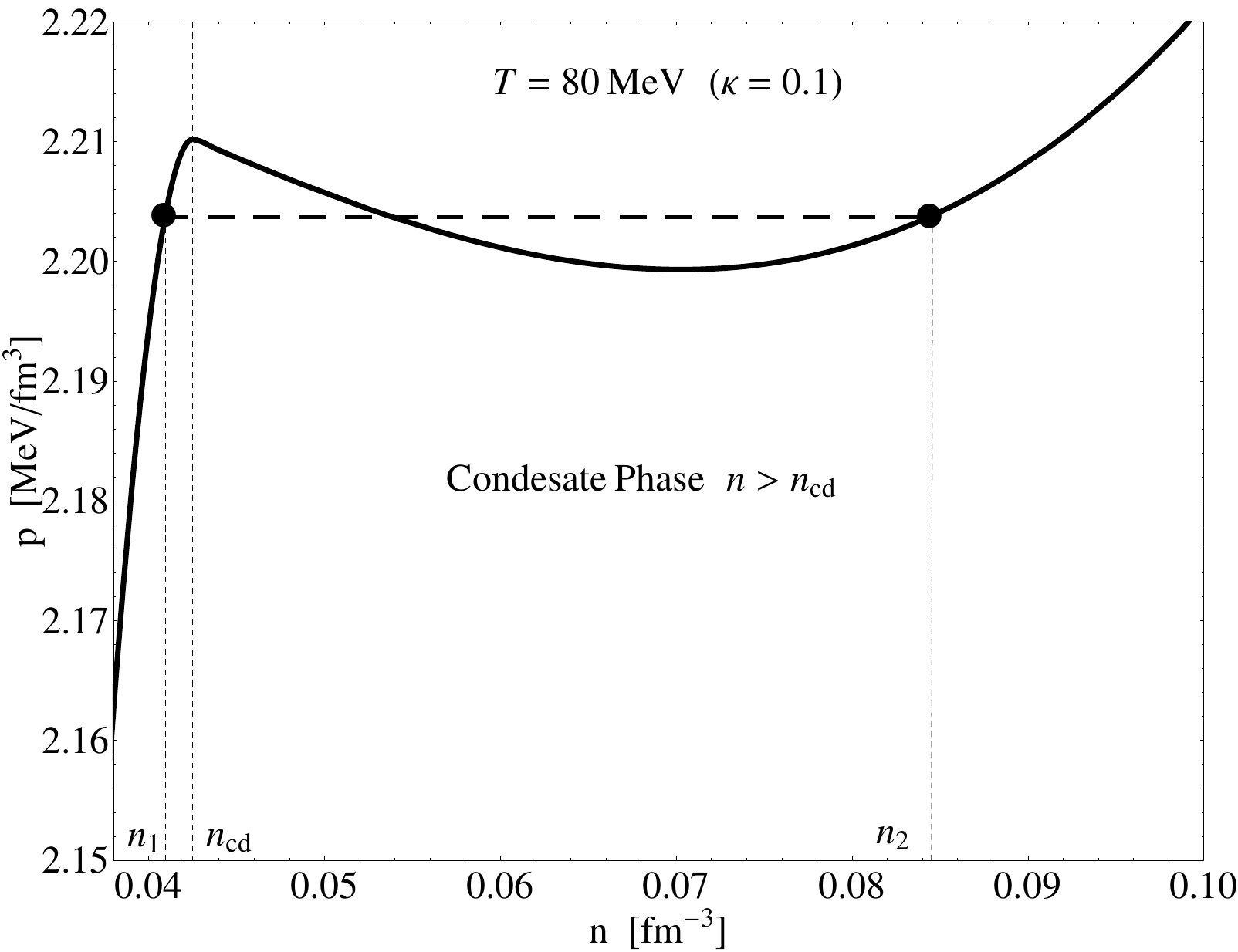}
\includegraphics[width=0.24\textwidth]{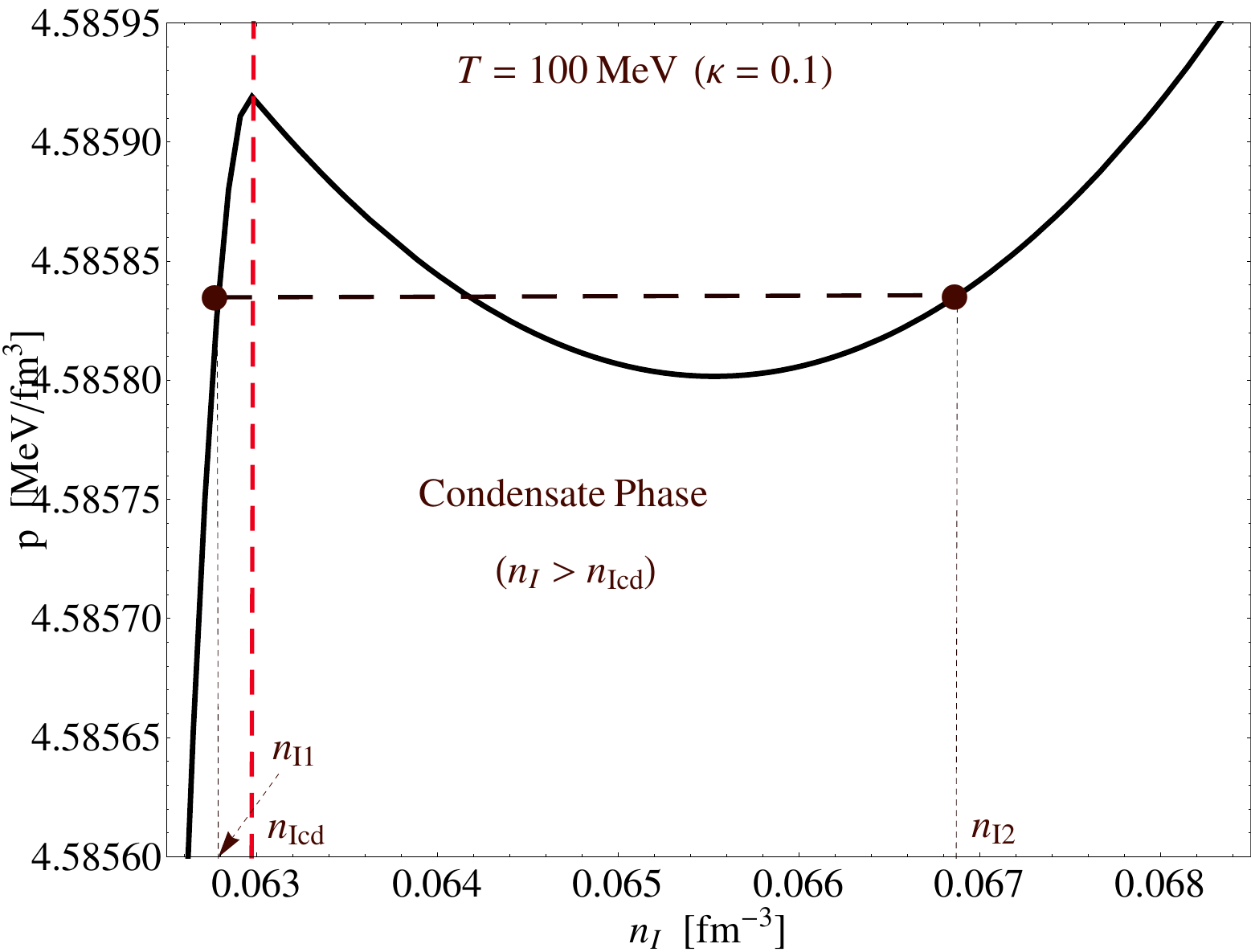}
\includegraphics[width=0.24\textwidth]{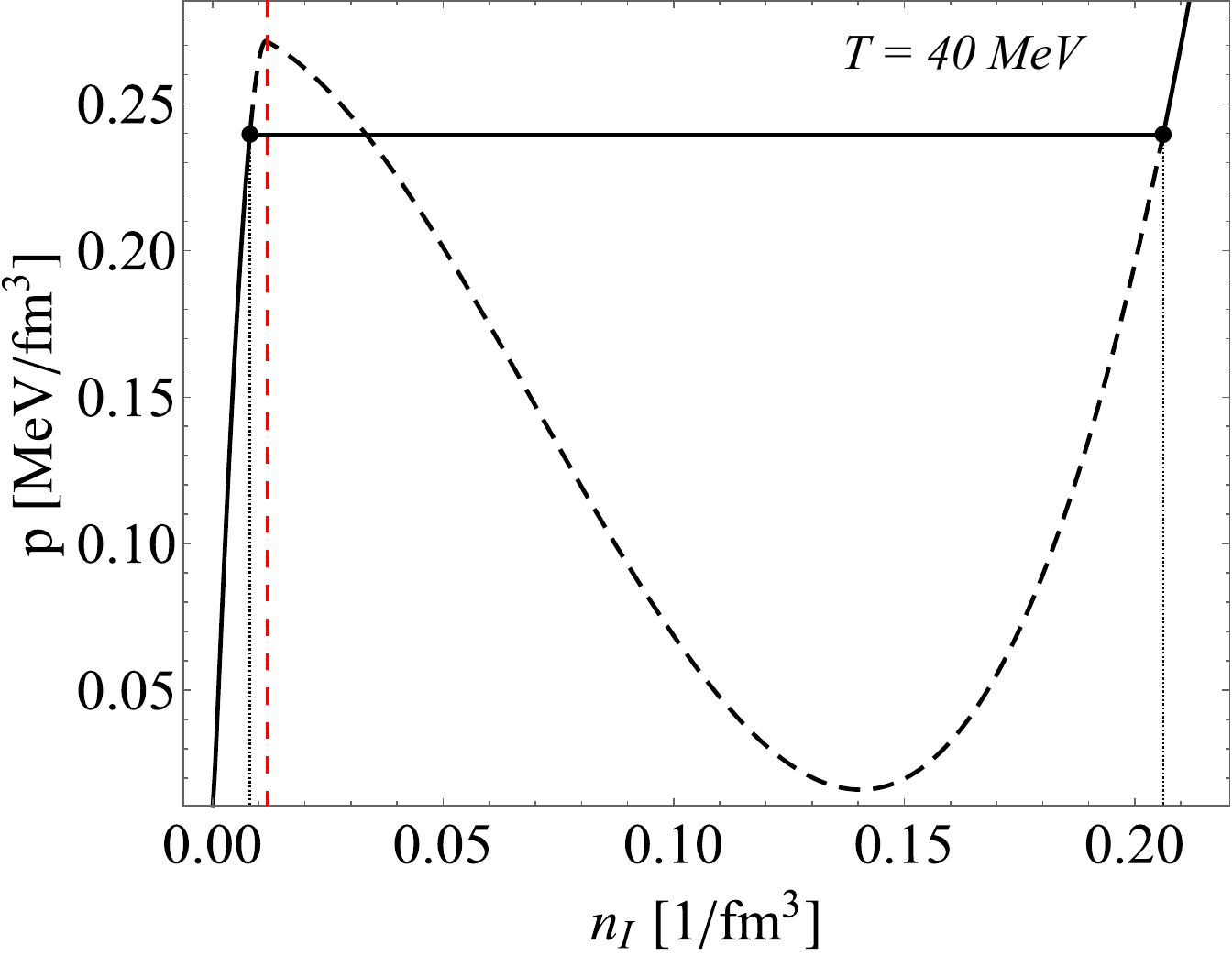}
\includegraphics[width=0.24\textwidth]{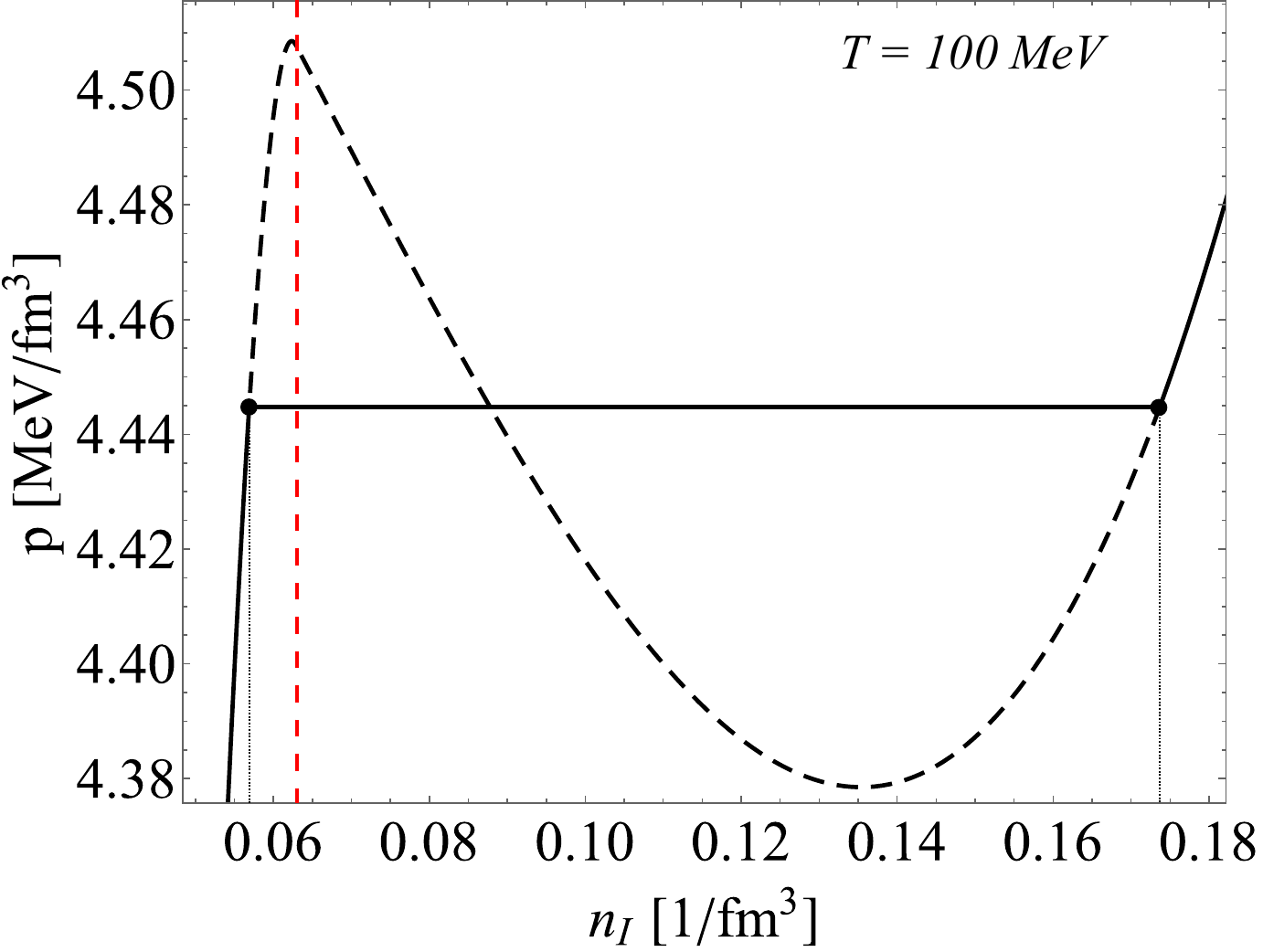}
\caption{
Dependence of pressure on the isospin density $n_I$ in the interacting
$\pi^+$-$\pi^-$ pion system within the framework of the mean-field model:
in the two left panels at $\kappa = 0.1$ and in the two right panels at
$\kappa = 0.2$.
}
\label{fig:p-vs-nI}
\end{figure}

The graphical examples of the application of this algorithm are depicted in
Figs.~\ref{fig:p-vs-invnI}-\ref{fig:p-vs-nI} for the set of temperatures
$T = 40,\, 80,\, 100$~MeV and two variations of the parameter
$\kappa = 0.1$ and $\kappa = 0.2$.
%
%

\section{Derivative of the pressure in the condensate phase}
\label{sec:deriv-pressure-cond}

Here, we present the details of calculating the pressure derivative
in the condensate phase.
As mentioned earlier, for the pressure in the interaction system we have
$p = p_{\rm kin} + P_{\rm ex}$.
For the derivative of the kinetic contribution, we obtain
\begin{equation}
\frac{\partial p_{\rm kin}(T,n_I)}{\partial  n_I} \,
=\, \frac{\partial p^{(-)}_{\rm kin}(T,n_I)}{\partial  n_I} \,
+\, \frac{\partial p^{(+)}_{\rm kin}(T,n_I)}{\partial  n_I} \,,
\qquad
\frac{\partial p^{(-)}_{\rm kin}(T,n_I)}{\partial  n_I} \,=\, 0 \,.
\label{eq:pkin-1}
\end{equation}
It follows from Eq.~(\ref{eq:d16}):
\begin{eqnarray}
p^{(+)}_{\rm kin}(T,n_I) = -\, T \int  \frac{d^3k}{(2\pi)^3} \,
\ln{ \left[ 1 - \exp{ \left( -\frac{\sqrt{m^2+{\bf k}^2} + 2U(n) + m }{T}\right)} \right] } \,,
\label{eq:pkin-3}
\end{eqnarray}
We calculate the derivative of this partial pressure:
\begin{equation}
\frac{\partial p^{(+)}_{\rm kin}(T,n_I)}{\partial  n_I} \,
=\, - \int  \frac{d^3k}{(2\pi)^3} \,
\frac{1}{\exp{ \left( \frac{\sqrt{m^2+{\bf k}^2} + 2U(n) + m }{T}\right)} - 1 } \,
2\frac{\partial U(n)}{\partial  n} \, \frac{\partial n}{\partial  n_I}  \,.
\label{eq:pkin-4}
\end{equation}
Hence, we get
\begin{equation}
\frac{\partial p_{\rm kin}(T,n_I)}{\partial  n_I} \,
=\, - 2 n^{(+)}(T,n_I) \frac{\partial U(n)}{\partial  n} \, \frac{\partial n}{\partial  n_I}  \,.
\label{eq:pkin-5}
\end{equation}
Now, using Eq.~(\ref{eq:dif-relation}), we can write
\begin{equation}
\frac{\partial p(T,n_I)}{\partial  n_I} \,
= \, - 2 n^{(+)}(T,n_I) \frac{\partial U(n)}{\partial  n} \, \frac{\partial n}{\partial  n_I} \,
+\, \frac{\partial P_{\rm ex}(n)}{\partial  n} \, \frac{\partial n}{\partial  n_I} \,
=\, \left( 1 - \frac{2 n^{(+)}(T,n_I)}{n(T,n_I)}  \right)
\, \frac{\partial P_{\rm ex}(n)}{\partial  n} \, \frac{\partial n}{\partial  n_I} \,
=\, 0 \,.
\label{eq:pkin-6}
\end{equation}
Taking into account $n = 2 n^{(+)}(T,n_I) + n_I$, we obtain that
$2 n^{(+)}/n < 1$ for finite isospin densities, $n_I > 0$.
Thus, for a positive value of the bracket $(1 - 2 n^{(+)}/n) > 0$, equation
(\ref{eq:pkin-6}) leads to
\begin{equation}
\frac{\partial P_{\rm ex}(n)}{\partial  n} \,=\, 0 \,.
\label{eq:pkin-7}
\end{equation}

\end{document}